\documentclass[usenatbib]{mnras}
\usepackage[utf8]{inputenc}
\usepackage{aas_macros}
\usepackage{amsfonts}
\usepackage{amsmath}
\usepackage{bm}
\usepackage{amssymb}
\usepackage{graphicx}
\usepackage[dvipsnames]{xcolor}
\usepackage{hyperref}
\hypersetup{colorlinks=true,allcolors=teal}

\defcitealias{ps22}{PS22}

\newcommand{\p}{\ensuremath{\partial}}

\newcommand{\xx}{\ensuremath{\mathbf{x}}}
\renewcommand{\aa}{\ensuremath{\mathbf{a}}}
\renewcommand{\aap}[1]{\ensuremath{\mathbf{a}_{\rm (p)#1}}}

\newcommand{\Mpch}{\ensuremath{h^{-1}{\rm Mpc}}}
\newcommand{\hMpc}{\ensuremath{h\,{\rm Mpc}^{-1}}}

\newcommand{\avg}[1]{\ensuremath{\left\langle \,#1\, \right\rangle}}
\newcommand{\e}[1]{\ensuremath{{\rm e}^{#1}}}

\newcommand{\der}{\ensuremath{{\rm d}}}

\newcommand{\erf}[1]{\ensuremath{{\rm erf}\left(#1\right)}}

\newcommand{\eqn}[1]{equation~\eqref{#1}}
\newcommand{\eqns}[1]{equations~\eqref{#1}}
\newcommand{\ph}[1]{\phantom{#1}}

\newcommand{\be}{\begin{equation}}
\newcommand{\ee}{\end{equation}}
\newcommand{\Cal}[1]{\ensuremath{\mathcal{#1}}}

\newcommand{\xiell}[1]{\ensuremath{\xi_{\rm NL}^{(#1)}}}

\newcommand{\Dellsq}[1]{\ensuremath{\Delta^{(#1)2}_{\rm NL}}}
\newcommand{\sigv}{\ensuremath{\sigma_{\rm v}}}
\newcommand{\Pell}[1]{\ensuremath{\mathcal{P}_{#1}}}
\newcommand{\beff}[2]{\ensuremath{b_{\rm eff}^{(#1)#2}}}
\newcommand{\sigeff}[2]{\ensuremath{\sigma_{\rm eff}^{(#1)#2}}}

\newcommand{\tjred}[3]{\ensuremath{\begin{pmatrix}
#1 & #2 & #3 \\
0 & 0 & 0 
\end{pmatrix}}}
\newcommand{\tjs}[6]{\ensuremath{\left(\begin{smallmatrix}
#1 & #2 & #3 \\
#4 & #5 & #6 
\end{smallmatrix}\right)}}
\renewcommand{\ss}{\ensuremath{\mathbf{s}}}
\newcommand{\eeb}{\ensuremath{\mathbf{e}}}
\newcommand{\rr}{\ensuremath{\mathbf{r}}}
\newcommand{\kk}{\ensuremath{\mathbf{k}}}

\newcommand{\sfid}{\ensuremath{\sigma_{\rm fid}}}
\newcommand{\qbar}[2]{\ensuremath{\bar{q}^{(#1)}_{#2}}}
\newcommand{\Qbar}[2]{\ensuremath{\bar{Q}^{(#1)}_{#2}}}

\title[Agnostic RSD]{Model-agnostic cosmological constraints from the baryon acoustic oscillation feature in redshift space} 
\author[Paranjape \& Sheth]{
 Aseem Paranjape$^{1}$\thanks{E-mail: aseem@iucaa.in} \& 
 Ravi K. Sheth$^{2,3}$\thanks{E-mail: shethrk@physics.upenn.edu},
\\  
 $^1$ Inter-University Centre for Astronomy \& Astrophysics, Ganeshkhind, Post Bag 4, Pune 411007, India\\
 $^2$ Center for Particle Cosmology, University of Pennsylvania, 209 S. 33rd St., Philadelphia, PA 19104, USA\\
 $^3$ The Abdus Salam International Center for Theoretical Physics, Strada Costiera, 11, Trieste 34151, Italy      }

\begin{document}
\label{firstpage}
\pagerange{\pageref{firstpage}--\pageref{lastpage}}
\maketitle

\begin{abstract}
We develop a framework for self-consistently extracting cosmological information from the clustering of tracers in redshift space, \emph{without} relying on model-dependent templates to describe the baryon acoustic oscillation (BAO) feature. 
Our approach uses the recently proposed Laguerre reconstruction technique for the BAO feature and its linear point $r_{\rm LP}$, and substantially extends it to simultaneously model the multipoles $\ell=0,2,4$ of the anisotropic galaxy 2-point correlation function (2pcf). 
The approach is `model-agnostic': it assumes that the non-linear growth of structure smears the BAO feature by an approximately Gaussian kernel with a smearing scale $\sigma_{\rm v}$, but does not assume any fiducial cosmology for describing the shape of the feature itself. 
Using mock observations for two realistic survey configurations assuming $\Lambda$ cold dark matter ($\Lambda$CDM), combined with Bayesian parameter inference, we show that the linear point $r_{\rm LP}$ and smearing scale $\sigma_{\rm v}$ can be accurately recovered by our method in both existing and upcoming surveys. 
The precision of the recovery of $r_{\rm LP}$ is always better than $1\%$, while $\sigma_{\rm v}$ can be recovered with $\lesssim10\%$ uncertainty provided the linear galaxy bias $b$ is separately constrained, e.g., using weak lensing observations. 
Our method is also sensitive to the linear growth rate $f$, albeit with larger uncertainties and systematic errors, especially for upcoming surveys such as DESI. 
We discuss how our model can be modified to improve the recovery of $f$, such that the resulting constraints on $\{f,\sigma_{\rm v},r_{\rm LP}\}$ can potentially be used as a test of cosmological models including and beyond $\Lambda$CDM. 
\end{abstract}

\begin{keywords}
cosmology: theory - methods: analytical, numerical
\end{keywords} 

\section{Introduction}
\label{sec:intro}
The baryon acoustic oscillation (BAO) feature in the distribution of galaxies and the inter-galactic medium has emerged as one of the most promising tools for the extraction of the cosmic distance scale (or standard ruler) related to the sound horizon at last scattering from multiple cosmological epochs \citep{eisenstein+05,cole+05,anderson+12,anderson+14,alam+17}. While the majority of current BAO analyses \citep[e.g.,][]{cuesta+16,beutler+17,blomqvist+19,dumasdesbourbouz+20,gil-marin+20,extractor2020,abbott+22} do this by fitting 2-point correlation function (2pcf) measurements near the BAO feature in configuration or Fourier space to templates inspired by a given cosmological model such as $\Lambda$ cold dark matter ($\Lambda$CDM), recent work has also shown how this extraction can be done in model-independent frameworks \citep{LP2016,LPmocks,LPboss,nsz21a,nsz21b}.

Among the latter, the use of Laguerre functions has been shown to be physically well-motivated \citep{nsz21a} in cosmological models -- such as $\Lambda$CDM -- in which gravitationally driven bulk flows smear the BAO feature \citep{bharadwaj96,cs06b} with an approximately Gaussian kernel of smearing scale \sigv. In this Laguerre reconstruction exercise, the `linear point' distance scale $r_{\rm LP}$ \citep{LP2016} can be extracted from measurements of the non-linear galaxy 2pcf \emph{without} relying on any cosmological model for a template \citep[e.g.,][]{gil-marin+20,hzs23} or physical reconstruction of galaxy positions \citep{esss07,padmanabhan+12}. Although the smearing scale \sigv\ is now a parameter in the problem, it is a single number as compared to a parametrised template shape, which makes the Laguerre reconstruction approach very appealing. However, it is not easy to disentangle the effects of \sigv\ in this approach from those of the coefficients \aa\ of the basis of Laguerre functions. While there has been some discussion regarding constraining \sigv\ simultaneously with these coefficients \citep{nsz22}, this has relied on approximations of the joint \emph{a posteriori} distribution $p(\aa,\sigv)$, and other analyses have typically assumed fixed values of \sigv\ \citep{nsz21a,nsz21b,ps22}.

In the present work, we show that measurements of the anisotropies in the observed galaxy 2pcf in redshift space can be leveraged to break the degeneracy between the basis coefficients \aa\ and the smearing scale \sigv, without any assumptions regarding their joint distribution. We will also argue that, as a bonus, the same analysis can be used in principle to constrain the logarithmic growth rate $f=\der\ln D/\der\ln a$, $D(a)$ being the linear theory growth factor. Our argument starts by exploiting the fact, noticed by \citet{nsz21b} and also used by \citet[][hereafter, PS22]{ps22}, that the monopole of the redshift space 2pcf can be treated identically to the real space 2pcf as regards the Laguerre reconstruction framework, with the replacement of linear halo bias $b$ and smearing scale \sigv\  with appropriately modified expressions that depend on $f$. We generalise this result to the multipoles \xiell{\ell} of the non-linear 2pcf $\xi_{\rm NL}$ for $\ell=0,2,4$ and show that the dependence of the resulting model on \sigv\ and $f$ can, in principle, allow us to constrain these  parameters simultaneously with the shape of the linear theory 2pcf. 

The resulting constraints on the cosmological variables $\{f,\sigv,r_{\rm LP}\}$ are model-agnostic. They do not explicitly assume a $\Lambda$CDM (or any other) cosmology at any stage during the fitting procedure. Instead, they assume that the Zel'dovich approximation \citep{Zeldovich70} is sufficiently accurate on the scales of interest, so that
the BAO feature is approximately smoothed by a Gaussian kernel, as mentioned above. This has interesting implications for testing cosmological models, including and beyond $\Lambda$CDM, which we briefly discuss. In this work, we present results based on some simplifying assumptions, the most important being that of scale-independent linear bias and the neglect of so-called `mode coupling' terms in galaxy power spectra. While these assumptions turn out to be acceptable for analysing current surveys, we show that they would be insufficient for upcoming surveys, and we discuss some avenues for improving the framework.

The paper is organised as follows. In section~\ref{sec:streaming}, we describe our main Zel'dovich smearing approximation and its effect on the 2pcf multipoles, including a discussion of the accuracy of the approximation. In section~\ref{sec:recon}, we show how the multipoles in this approximation can be modelled in the Laguerre reconstruction framework without any reference to the shape of the 2pcf in any fiducial cosmology. We test our framework on mock data, whose generation and analysis we describe in section~\ref{sec:analysis}, with results presented in section~\ref{sec:results} along with a discussion of shortcomings in the current model and possible ways forward. We summarise and conclude in section~\ref{sec:conclude}. Wherever needed, we assume the Baryon Oscillation Spectroscopic Survey (BOSS) Final Year flat $\Lambda$CDM cosmology \citep{alam+17} with parameters $\{\Omega_{\rm m},\Omega_{\rm b},h,n_{\rm s},\sigma_8\}=\{0.31,0.04814,0.676,0.97,0.8\}$ and use a linear theory transfer function calculated with the code \textsc{class} \citep{class-I,class-II}.\footnote{\url{http://class-code.net}}

\section{Zel'dovich smearing approximation}
\label{sec:streaming}
In this section, we describe the analytical framework, based on an anisotropic smearing approximation in $\Lambda$CDM, that motivates our model-agnostic reconstruction setup of section~\ref{sec:recon}.

Throughout, we will use boldface symbols to denote vectors and tensors, and symbols with carets to denote unit vectors, so that, e.g., $\kk=k\,\hat{k}$.
We will use \Pell{\ell} to denote the Legendre polynomials, defined as
\be
\Pell{\ell}(\mu) = \frac{1}{2^\ell\ell!}\frac{\der^\ell}{\der\mu^\ell}\left(\mu^2-1\right)^\ell\,;\quad \ell=0,1,\ldots\,,
\label{eq:Legendre-def}
\ee
where $-1\leq\mu\leq1$, in terms of which the multipoles $h_\ell$ of some function $h(\mu)$ are defined as
\be
h_\ell = (2\ell+1)\int_{-1}^1\frac{\der\mu}{2}\,\Pell{\ell}(\mu)\,h(\mu)\,.
\label{eq:multipole-def}
\ee
We will also need the following integral relations obeyed by the Legendre polynomials,
\begin{align}
\int_{-1}^1\frac{\der\mu}{2}\,\Pell{\ell}(\mu)\,\Pell{\ell^\prime}(\mu) &= \frac{1}{(2\ell+1)}\,\delta_{\ell\ell^\prime}\,,
\label{eq:Legendre-orthog}\\
\int_{-1}^1\frac{\der\mu}{2}\,\Pell{\ell}(\mu)\,\Pell{\ell^\prime}(\mu)\,\Pell{\ell^{\prime\prime}}(\mu) &= 
\tjred{\ell}{\ell^\prime}{\ell^{\prime\prime}}^2
\,,
\label{eq:Legendre-3j}
\end{align}
where $\tjs{j_1}{j_2}{j_3}{m_1}{m_2}{m_3}$ is a Wigner $3j$ symbol \citep{Wigner1993}. Finally, we will always assume the plane parallel approximation with line-of-sight direction $\hat{n}$, and use the symbol $\mu$ with subscripts to indicate the cosine of the angle between a given vector and $\hat{n}$, e.g.,
\be
\mu_s \equiv \hat{s}\cdot\hat{n}\,;\quad \mu_k\equiv\hat{k}\cdot\hat{n}\,.
\label{eq:mu-def}
\ee

\subsection{Smearing due to displacements}
If $\xi_{\rm L}(\rr)$ and $\xi_{\rm NL}(\rr)$ are, respectively, the linear and non-linearly evolved galaxy 2pcf in real space, and we focus on large separations $r$ close to the BAO feature, then the effect of bulk flows is to smear the linear 2pcf according to \citep{bharadwaj96,cs06b,cs08}
\be
\xi_{\rm NL}(\rr) \approx \int\der^3r^\prime\,\xi_{\rm L}(\rr^\prime)\,\Cal{N}\left(\rr-\rr^\prime;\,2\sigv^2\,\boldsymbol{1}\right)\,,
\label{eq:smearing-realspace}
\ee
where $\Cal{N}(\xx;\mathbf{\Sigma})$ denotes a 3-dimensional Gaussian distribution in \xx\ having zero mean and covariance matrix $\mathbf{\Sigma}$, with $\boldsymbol{1}$ indicating the identity matrix, we defined the linear theory 1-dimensional, single-particle velocity dispersion \sigv\ (in units of comoving length) using
\be
\sigv^2\equiv \frac13\int\der\ln k\,k^{-2}\,\Delta^2_{\rm lin}(k)\,,
\label{eq:sigv-def}
\ee
with $\Delta^2_{\rm lin}(k)=k^3P_{\rm lin}(k)/(2\pi^2)$ being the dimensionless linear theory matter power spectrum (whose redshift dependence we suppress), and we ignored the effects of mode coupling and scale-dependent bias in \eqn{eq:smearing-realspace}. They can be included following, e.g., \citet{peaksRSD}, and we discuss the systematic biases introduced by their neglect in section~\ref{subsec:caveats}. The assumption of scale-independent bias implies
\be
\xi_{\rm L}(\rr)=b^2\xi_{\rm lin}(r)=b^2\int\der\ln k\,\Delta^2_{\rm lin}(k)\,j_0(kr)\,,
\label{eq:xiL-real}
\ee
where $b$ is the linear, scale-independent galaxy bias and $j_\ell(x)$ denotes the spherical Bessel function of order $\ell$.

In redshift space, the covariance matrix of the Gaussian kernel in \eqn{eq:smearing-realspace} becomes anisotropic: $2\sigv^2\,\boldsymbol{1}\to\mathbf{\Sigma}$, with variances $\sigma_\parallel^2$ and $\sigma_\perp^2$ along and perpendicular to the line-of-sight, respectively, given by \citep{ZeldovichRSD,peaksRSD,ppvv15}
\begin{align}
\sigma_\parallel^2 = 2\sigv^2\,(1+f)^2\,\quad {\rm and}\quad 
\sigma_\perp^2 = 2\sigv^2\,,
\label{eq:sigpar-sigperp}
\end{align}
which allows us to write the non-linearly evolved 2pcf in redshift space $\xi_{\rm NL}(\ss)$ in terms of the linear 2pcf in redshift space $\xi_{\rm L}(\ss)$ as
\be
\xi_{\rm NL}(\ss) \approx \int\der^3s^\prime\,\xi_{\rm L}(\ss^\prime)\,\Cal{N}\left(\ss-\ss^\prime;\,\mathbf{\Sigma}\right)\,,
\label{eq:smearing-redshiftspace}
\ee
with 
\be
\xi_{\rm L}(\ss) = \int\frac{\der^3k}{(2\pi)^3}\,\e{i\kk\cdot\ss}\,b^2\left(1+\beta\mu_k^2\right)^2\,P_{\rm lin}(k)\,,
\label{eq:xiL-redshift}
\ee
where $\beta\equiv f/b$ \citep{kaiser87}.
We can now find expressions for the multipoles $\xiell{\ell}(s)$ of $\xi_{\rm NL}(\ss)$ as follows:
\begin{align}
&(2\ell+1)^{-1}\,\xiell{\ell}(s) \notag\\
&= \int_{-1}^1\frac{\der\mu_s}{2}\,\Pell{\ell}(\mu_s)\xi_{\rm NL}(\ss) \notag\\
&= \int_{-1}^1\frac{\der\mu_s}{2}\,\Pell{\ell}(\mu_s)\int\der^3s^\prime\,\xi_{\rm L}(\ss^\prime)\int\frac{\der^3k}{(2\pi)^3}\,\e{i\kk\cdot(\ss-\ss^\prime)}\,\e{-\kk^{\rm T}\cdot\mathbf{\Sigma}\cdot\kk/2} \notag\\
&= \int\frac{\der k\,k^2}{2\pi^2}\int_{-1}^1\frac{\der\mu_k}{2}\,
b^2\,P_{\rm lin}(k)\left(1+\beta\mu_k^2\right)^2\,\e{-k^2\sigv^2\left[1+\mu_k^2f(f+2)\right]}\notag\\
&\ph{\int\frac{\der k\,k^2}{2\pi^2}\int_{-1}^1\frac{\der\mu_k}{2}\,b^2}
\times\int_{-1}^1\frac{\der\mu_s}{2}\,\Pell{\ell}(\mu_s)\int_0^{2\pi}\frac{\der\phi_k}{2\pi}\,\e{i\kk\cdot\ss}\,.
\label{eq:xiell-1}
\end{align}
The plane wave expansion 
\be
\e{i\kk\cdot\ss} = 4\pi\sum_{\ell^\prime=0}^\infty i^{\ell^\prime}\,j_{\ell^\prime}(ks)\sum_{m^\prime=-\ell^\prime}^{\ell^\prime}\,Y^{m^\prime}_{\ell^\prime}(\hat{k})\,Y^{m^\prime\ast}_{\ell^\prime}(\hat{s})
\ee
in terms of spherical harmonics
$Y^m_\ell$
then gives
\be
\int_0^{2\pi}\frac{\der\phi_k}{2\pi}\,\e{i\kk\cdot\ss} = \sum_{\ell^\prime=0}^\infty i^{\ell^\prime}\,j_{\ell^\prime}(ks)\,(2\ell^\prime+1)\Pell{\ell^\prime}(\mu_k)\Pell{\ell^\prime}(\mu_s)\,,
\ee
so that
\be
\int_{-1}^1\frac{\der\mu_s}{2}\,\Pell{\ell}(\mu_s)\int_0^{2\pi}\frac{\der\phi_k}{2\pi}\,\e{i\kk\cdot\ss} = i^\ell\,j_\ell(ks)\,\Pell{\ell}(\mu_k)\,.
\ee
Plugging this into \eqn{eq:xiell-1} gives us
\be
\xiell{\ell}(s) = i^\ell\int\der\ln k\,j_\ell(ks)\,\Dellsq{\ell}(k)\,,
\label{eq:xiell-2}
\ee
where we defined
\begin{align}
\Dellsq{\ell}(k) &\equiv \Delta_{\rm lin}^2(k)\,b^2\,\e{-k^2\sigv^2}\,(2\ell+1)\notag\\
&\ph{b^2}
\times\int_{-1}^1\frac{\der\mu_k}{2}\,\Pell{\ell}(\mu_k)\left(1+\beta\mu_k^2\right)^2\e{-K^2\mu_k^2}\,,
\label{eq:Delta(ell)2-def}
\end{align}
with 
\be
K^2 \equiv k^2\sigv^2\,f(f+2)\,.
\label{eq:K2-def}
\ee
The integral over $\mu_k$ can be done analytically (see Appendix~\ref{sec:Cgp}); for any $\ell$, it is the sum of terms which multiply $\erf{K}/K$ and others which multiply $\exp(-K^2)$, but these are not very illuminating.  If expanded as a Taylor series (in $K^2$), then it has terms of alternating sign, so any truncation must be done with care.  We discuss this below.
\emph{Hereon, we will only be interested in the multipoles $\ell=0,2,4$, and all summations $\sum_\ell$ over $\ell$ will be restricted to these three values, unless explicitly stated otherwise.}

\begin{figure*}
\centering
\includegraphics[width=0.42\textwidth,trim=7 8 6 4,clip]{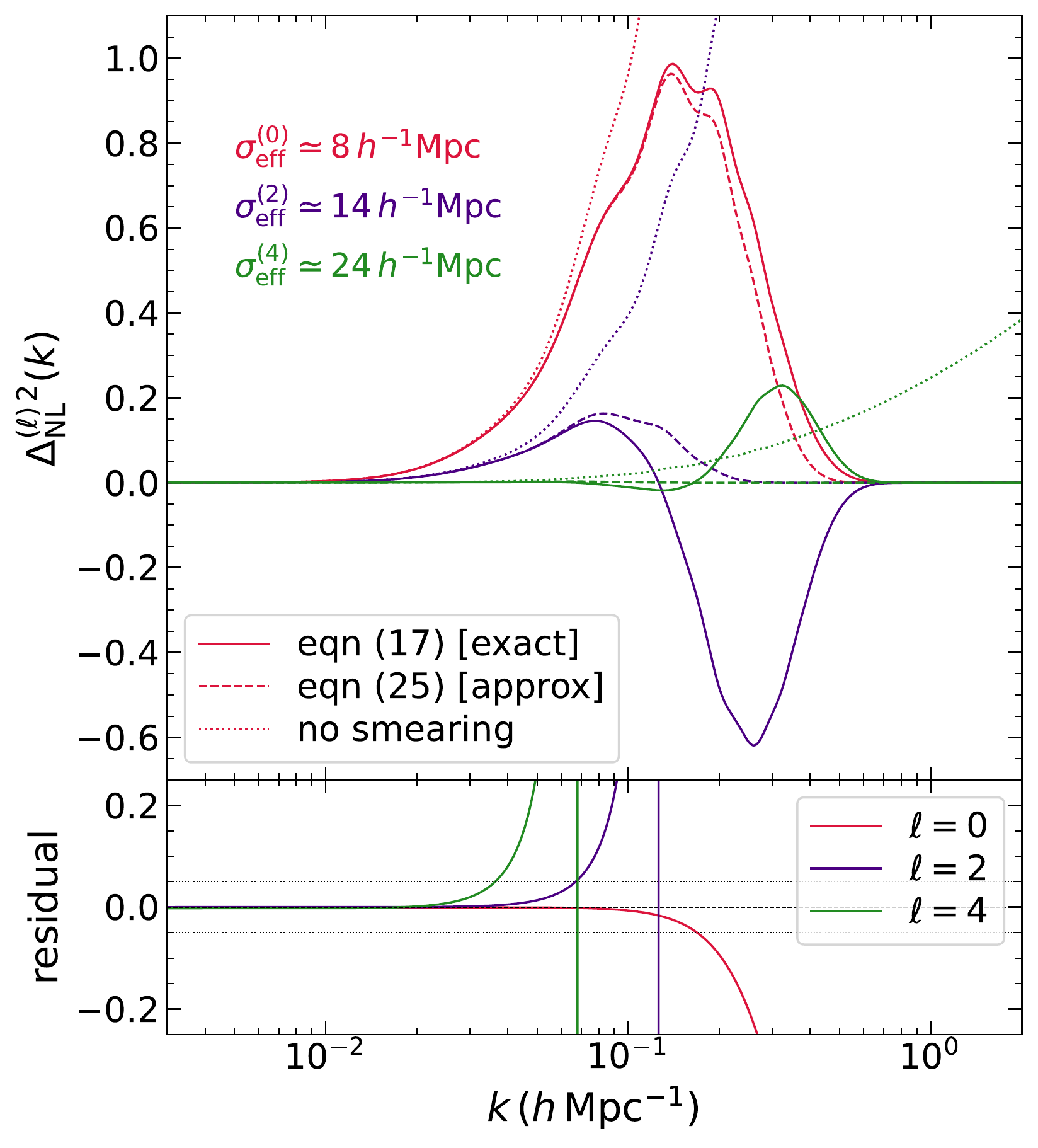}
\includegraphics[width=0.42\textwidth,trim=5 8 6 6,clip]{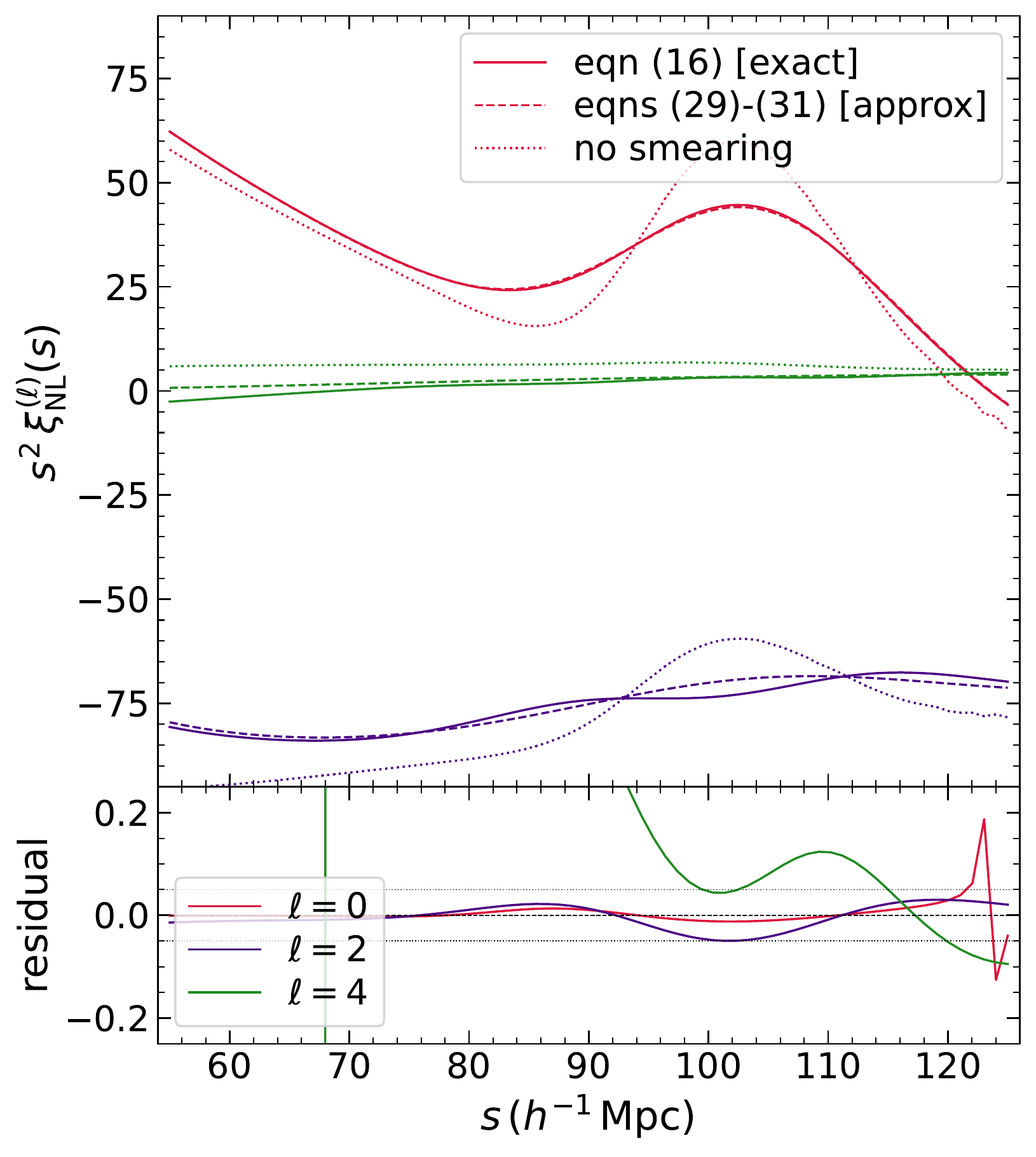}
\caption{Multipoles of the non-linearly evolved redshift space dimensionless power spectrum \emph{(left panel)} and 2pcf \emph{(right panel)} in the DESI LRG configuration (Table~\ref{tab:toy_samples}). 
In each \emph{upper} panel, solid curves show the exact result assuming scale-independent bias and ignoring mode coupling, dashed curves show the Zel'dovich smearing approximation, and dotted curves show the limit of no smearing; it is the obvious differences between these and the other curves which motivates Laguerre reconstruction. The text labels in the \emph{left upper panel} give the estimated input values of \sigeff{\ell}{} (equation~\ref{eq:sigeff-def}) for this configuration. The solid curves in the \emph{lower panels} show the corresponding residuals (i.e., approximate/exact - 1), while the horizontal dotted lines indicate $\pm5\%$ deviations. See text for a discussion.}
\label{fig:diagnostic}
\end{figure*}

To simplify \eqn{eq:Delta(ell)2-def}, we start by noting the identity 
\be
\left(1+\beta\mu_k^2\right)^2 = \sum_\ell\,\chi_\ell(\beta)\,\Pell{\ell}(\mu_k)\,, 
\label{eq:Kaiser_multipoles}
\ee
where, following \cite{hamilton92}, we defined
\begin{align}
\chi_0(\beta) &\equiv 1 + \frac{2\beta}{3} + \frac{\beta^2}{5}\,,\notag\\
\chi_2(\beta) &\equiv 4\beta\left(\frac{1}{3} + \frac{\beta}{7}\right)\,,\notag\\
\chi_4(\beta) &\equiv \frac{8\beta^2}{35}\,.
\label{eq:chiell-def}
\end{align}
Next, we use the fact that we are only interested in redshift space separations close to the BAO feature, so that $s\sim100\Mpch$, while the bulk flow smearing scale $\sigv\sim10\Mpch$, so that $\sigv/s\ll1$. Since the Bessel functions in \eqn{eq:xiell-2} effectively restrict the integral over $k$ to values $k\lesssim 1/s$, in \eqn{eq:Delta(ell)2-def} we can safely assume $K^2\ll1$, so that $\e{-K^2\mu_k^2} = 1-K^2\mu_k^2 + \Cal{O}(K^4)$ under the integral over $\mu_k$ (since $|\mu_k|\leq1$). Finally, we use \eqns{eq:Legendre-orthog} and \eqref{eq:Legendre-3j} to obtain the identity
\begin{align}
&(2\ell+1)\int_{-1}^1\frac{\der\mu}{2}\,\Pell{\ell}(\mu)\,\Pell{\ell^\prime}(\mu)\,\mu^2 \notag\\
&\ph{(2\ell+1)}
=\frac13\left[\delta_{\ell,\ell^\prime} + 2(2\ell+1)\tjred{\ell}{\ell^\prime}{2}^2\right]\,.
\end{align}
Putting all this together, the integral in \eqn{eq:Delta(ell)2-def} becomes
\begin{align}
&(2\ell+1)\int_{-1}^1\frac{\der\mu_k}{2}\,\Pell{\ell}(\mu_k)\left(1+\beta\mu_k^2\right)^2\e{-K^2\mu_k^2}\notag\\
&\ph{2\ell}
= \chi_\ell(\beta)\bigg[1 - \frac{K^2}{3}\left\{1 + \frac{2(2\ell+1)}{\chi_\ell(\beta)}\sum_{\ell^\prime}\chi_{\ell^\prime}(\beta)\tjred{\ell}{\ell^\prime}{2}^2\right\}\notag\\
&\ph{2\ell\sum_{\ell^\prime}\chi_{\ell^\prime}(\beta)\delta_{\ell\ell^\prime}}
+\Cal{O}(K^4)\bigg]\notag\\
&\ph{2\ell}
= \chi_\ell(\beta)\exp\left[-\frac{K^2}{3}\left(1+\frac{2(2\ell+1)}{\chi_\ell(\beta)}\sum_{\ell^\prime}\Cal{C}_{\ell\ell^{\prime}}\chi_{\ell^\prime}(\beta)\right)\right]\notag\\
&\ph{\chi_\ell(\beta)\exp[]}
\times\left(1+\Cal{O}(K^4)\right)\,,
\label{eq:expintegral}
\end{align}
where, in the last equality, we defined the symmetric matrix $\Cal{C}_{\ell\ell^\prime}$,
\be
\Cal{C}_{\ell\ell^\prime} \equiv \tjred{\ell}{\ell^\prime}{2}^2\,,
\label{eq:Cmatrix-def}
\ee
with $\ell$ and $\ell^\prime$ taking values $0,2,4$. Evaluating $\Cal{C}_{\ell\ell^\prime}$ for these values using the definition of the Wigner $3j$ symbols gives 
\be
\Cal{C}_{\ell\ell^\prime} = 
\begin{pmatrix}
\Cal{C}_{00} & \Cal{C}_{02} & \Cal{C}_{04} \\
\Cal{C}_{20} & \Cal{C}_{22} & \Cal{C}_{24} \\
\Cal{C}_{40} & \Cal{C}_{42} & \Cal{C}_{44} \end{pmatrix}
=
\begin{pmatrix}
0 & 1/5 & 0 \\
1/5 & 2/35 & 2/35 \\
0 & 2/35 & 20/693 
\end{pmatrix}\,.
\label{eq:Cmatrix-values}
\ee
The expression for $\Dellsq{\ell}(k)$ in  \eqn{eq:Delta(ell)2-def} simplifies to
\be
\Dellsq{\ell}(k) = \Delta_{\rm lin}^2(k)\,\beff{\ell}{2}\,\e{-k^2\sigeff{\ell}{2}/2}\left(1+\Cal{O}(K^4)\right)\,,
\label{eq:Delta(ell)2}
\ee
where we defined the effective bias \beff{\ell}{} and effective smearing scale \sigeff{\ell}{} using
\begin{align}
\beff{\ell}{2} &\equiv b^2\,\chi_{\ell}(\beta) \,,
\label{eq:beff-def}\\
\sigeff{\ell}{2} &\equiv 2\,\sigv^2\left[1 + \frac{f(f+2)}{3}\left(1+\frac{2(2\ell+1)}{\chi_{\ell}(\beta)}\sum_{\ell^\prime}\Cal{C}_{\ell\ell^\prime}\,\chi_{\ell^\prime}(\beta)\right)\right]\,.
\label{eq:sigeff-def}
\end{align}
In effect, we have dealt with the fact that $\exp(-K^2\mu_k^2)$ has alternating signs by evaluating the integral to lowest order in $K^2$, and then used the result to define an effective smearing scale.  As a result, upto terms of order $\Cal{O}(K^4)$, the Fourier transform of $\Dellsq{\ell}(k)/k^3$ is explicitly an \emph{isotropic} Gaussian smearing of the linear 2pcf with the replacements $b\to \beff{\ell}{}$ and $2\sigv^2\to\sigeff{\ell}{2}$ in \eqns{eq:smearing-realspace} and~\eqref{eq:xiL-real}. For $\ell=0$, this recovers the derivation of $\xiell{0}(s)$ in \citet{nsz21b}. For $\ell=2,4$, however, the presence of $j_2$ and $j_4$ in \eqref{eq:xiell-2} means that the Fourier transform of $\Dellsq{\ell}(k)/k^3$ does not directly appear in \xiell{2} and \xiell{4}.

To proceed further, we manipulate the Bessel functions to simplify the expressions for \xiell{\ell} as
described in Appendix~\ref{app:streamingapprox}. As we show there, it is useful to define the quantity
\be
\xi_0(s|\sigma) \equiv \int\der\ln k\,j_0(ks)\,\Delta_{\rm lin}^2(k)\,\e{-k^2\sigma^2/2}\,,
\label{eq:xi0(s|sigma)}
\ee
which is exactly of the form for \xiell{0}, except that the Gaussian smearing uses the generic scale $\sigma$. In terms of this, the smearing approximation reduces to 
\begin{align}
\xiell{0}(s) &= \beff{0}{2}\,\xi_0(s|\sigeff{0}{})\,, \label{eq:xi0NL-final}\\
\xiell{2}(s) &= \beff{2}{2}\left[\xi_0(s|\sigeff{2}{}) - \bar{\xi}_0(s|\sigeff{2}{})\right]\,, \label{eq:xi2NL-final}\\
\xiell{4}(s) &= \beff{4}{2}\left[\xi_0(s|\sigeff{4}{}) + \frac52\bar{\xi}_0(s|\sigeff{4}{}) - \frac72\bar{\bar{\xi}}_0(s|\sigeff{4}{})\right]\,,
\label{eq:xi4NL-final}
\end{align}
where ${\bar \xi}_0(s|\sigma)$ and ${\bar {\bar \xi}}_0(s|\sigma)$ are defined in \eqn{eq:xibar-def} and \eqn{eq:xibarbar-def}, respectively. Note especially the appearance of \sigeff{\ell}{} in the expression for \xiell{\ell}, for each $\ell$. We refer to \eqns{eq:xi0NL-final}-\eqref{eq:xi4NL-final} as our \emph{Zel'dovich smearing approximation} in what follows.  In the limit of no smearing ($\sigeff{l}{}\to 0$) these expressions reduce to equations~(6-8) of \cite{hamilton92}.

\subsection{Accuracy of the smearing approximation}
\label{subsec:modelaccuracy}
If the effects of mode coupling and scale-dependent bias can be ignored, the key step that allows us to approximate \eqn{eq:xiell-2} by \eqns{eq:xi0NL-final}-\eqref{eq:xi4NL-final} (equivalently, equation~\ref{eq:Delta(ell)2-def} by equation~\ref{eq:Delta(ell)2}) is the assumption that terms of order $\Cal{O}(K^4)$ in \eqn{eq:Delta(ell)2} can be approximately resummed into an exponential smearing term. There are several aspects of this approximation that bear some discussion.

The \emph{left panel} of Fig.~\ref{fig:diagnostic} compares the full integral expression \eqref{eq:Delta(ell)2-def} for $\Dellsq{\ell}$ (solid) with the no smearing limit (dotted) and our Zel'dovich smearing approximation \eqref{eq:Delta(ell)2} (dashed) in the DESI LRG configuration described later, for $\ell=0,2$ and $4$.  The dotted curves have the same shape but different amplitudes ($\propto \beff{\ell}{2}$).  At each $\ell$, they are similar to the solid curves only at very small $k$, but they are otherwise {\em very} different:  this is why some accounting for the smearing must be made.  The dashed curves show that our approximation \eqref{eq:Delta(ell)2} fares much better.  It loses fidelity at increasingly smaller $k$ as $\ell$ increases, with $\sim5\%$ inaccuracies being reached at $k\simeq 0.17, 0.07, 0.035\,\hMpc$ for $\ell=0,2,4$, respectively. This is essentially a consequence of the fact that, having repackaged the $\Cal{O}(K^2)$ term into an exponential, the resulting \sigeff{\ell}{} \emph{increases} with $\ell$ (see the text labels in the \emph{left panel} of Fig.~\ref{fig:diagnostic}). For comparison, $\sqrt{2}\sigv\simeq5.6\Mpch$ for this choice of parameters.  In addition, this repackaging fails badly at large $k$, since it cannot reproduce the fact that, for $\ell=2$ and 4, the `exact' result changes sign.  (Note, however, that the `exact' result ignores the effects of mode coupling and scale-dependent bias, both of which likely matter at large $k$.)

To see how this affects the BAO feature in configuration space, the \emph{right panel} of Fig.~\ref{fig:diagnostic} compares the full integral \eqref{eq:xiell-2} for $\xiell{\ell}(s)$ (solid) with the  no-smearing limit (dotted), and the smearing approximation in \eqns{eq:xi0NL-final}-\eqref{eq:xi4NL-final} (dashed). The latter was numerically evaluated using the identities \eqref{eq:xi0bars-integrals}. In this case, the no-smearing limit is again quite bad for all $\ell$, whereas our smearing approximation for \xiell{0} is accurate at better than $\sim1\%$ over nearly the entire range of scales of interest, failing in a relative sense only in the near vicinity of the zero-crossing of the function ($s\sim123\Mpch$). The approximation for \xiell{2} is accurate at better than $\sim5\%$ over the entire range. While this level of agreement is remarkable, it is also clear that the approximation does not capture the oscillatory features of wavelength $\lambda_{\rm osc}\sim30\Mpch$ seen near the BAO feature in the exact integral. These oscillations can be traced to the behaviour of the solid purple curve for $\Dellsq{2}(k)$ in the \emph{left panel}, which has a (negative) spike in power at $k\sim0.2\,\hMpc\approx2\pi/\lambda_{\rm osc}$. The smearing approximation (dashed purple) has $\sigeff{2}{}\simeq14\Mpch$, which understandably washes over the oscillations and cannot reproduce the change in sign of the $k$-space power. 
Finally, the approximation for \xiell{4}, although very close to the exact integral in an absolute sense, shows $>20\%$ deviations at $s<95\Mpch$. This is not surprising, considering that we saw $\gtrsim5\%$ inaccuracies in the corresponding $k$-space approximation at $k\gtrsim0.035\,\hMpc$ in the \emph{left panel}.\footnote{Similarly to $\Dellsq{2}(k)$, the exact expression for $\Dellsq{4}(k)$ also shows a spike near $k\sim0.3\hMpc$ in the \emph{left panel} of Fig.~\ref{fig:diagnostic}. Due to the lower magnitude of this feature, however, the corresponding oscillations of wavelength $\sim20\Mpch$ in \xiell{4} in the \emph{right panel} have a much smaller amplitude. These are, however, noticeable in the residual of $\xiell{4}(s)$ near $s\simeq110\Mpch$.}

This comparison suggests that, in practice, while the smearing approximation \eqref{eq:xi0NL-final} for \xiell{0} is expected to be very accurate near the BAO feature, the approximation for \xiell{2} is likely to introduce biases when the observational errors on \xiell{2} approach $\sim5\%\,\times\sqrt{N}$ for $N$ independent measurements. Similarly, for \xiell{4}, the increasing inaccuracy of the model at $s\lesssim90\Mpch$ makes it likely that accurate measurements of \xiell{4} could introduce biases in the inferred parameters. 

Another consequence of the $\ell$-dependence of \sigeff{\ell}{} is that the steepness of the increase with $\ell$ is especially pronounced for highly biased tracers, for which $\beta$ can be small. This is quite different from the $\ell=0$ behaviour, because $\chi_\ell(\beta)\to1$ in the denominator of the last term in \eqn{eq:sigeff-def} for $\ell=0$ but diverges for other $\ell$, if $\beta\to0$.  This can potentially complicate the Laguerre reconstruction described in section~\ref{sec:recon}, for which a useful rule of thumb is that the BAO feature should be about one smearing length away from the smallest and largest scales being modelled.

The level to which these inaccuracies in the model affect parameter recovery will depend on the details of the survey in question. We return to this point in section~\ref{sec:results} where we show results for two mock survey configurations.

\section{Multipole reconstruction}
\label{sec:recon}
If we model the real space linear theory 2pcf $\xi_{\rm lin}(r)$ in some range $r_{\rm min}\leq r\leq r_{\rm max}$ as the simple polynomial of degree $M-1$,
\be
b^2\xi_{\rm lin}(r) = \sum_{m=0}^{M-1}\frac{a_m}{m!}\left(\frac{r}{\sfid}\right)^m\,,
\label{eq:xilin-poly}
\ee
where \sfid\ is a fiducial scale used to non-dimensionalise the problem, then the Gaussian convolution $\xi_0(s|\sigma)$ becomes
\be
b^2\xi_0(s|\sigma) = \sum_{m=0}^{M-1}\frac{a_m}{m!}\left(\frac{\sigma}{\sfid}\right)^m\,\mu_m(s/\sigma)\,,
\label{eq:xi0-Lag}
\ee
in a corresponding range $s_{\rm min}\leq s\leq s_{\rm max}$, where the $\mu_m(x)$ are given by equation~(7) of \citet{nsz21a} in terms of generalised Laguerre functions.

\citetalias{ps22} exploited the fact that, when focusing on the redshift space monopole \xiell{0} and fixing the smearing scale $\sigma=\sigeff{0}{}$, the coefficients \aa\ give a representation of the \emph{un-smeared} real space linear 2pcf in some chosen range of scales. If we think of this as saying that the coefficients \aa\ describe the behaviour of (the Fourier transform of) $\Delta_{\rm lin}^2(k)/k^3$, then the derivation in section~\ref{sec:streaming} shows that it is only the value of the smearing scale in the factor $\sim\e{-k^2\sigma^2/2}$ which changes when considering different multipoles. This means that the \emph{same} polynomial coefficients \aa\ should give a good description of \emph{all} multipoles \xiell{\ell}, $\ell=0,2,4$, provided one appropriately modifies the smearing scale (i.e., uses $\sigma=\sigeff{\ell}{}$) and propagates the effect of volume averaging of the resulting Laguerre functions. In this case, $b$, $\beta$ and \sigv\ would be free parameters that must be jointly constrained with \aa. The fact that the polynomial description of the linear 2pcf is only valid over a finite range of scales $s_{\rm min}\leq s\leq s_{\rm max}$ can be accounted for by introducing three new free parameters, which would capture the (constant) integrals of $u^2\xi_0(u|\sigeff{2}{})$, $u^2\xi_0(u|\sigeff{4}{})$ and $u^4\xi_0(u|\sigeff{4}{})$ over the range $0\leq u\leq s_{\rm min}$.

In this section, we describe how to implement this approach. In addition to the linear point $r_{\rm LP}$, this approach can potentially constrain the cosmological parameters $\beta$ and \sigv\ \emph{without} assuming any fiducial cosmology. We will also discuss the role of the linear bias parameter $b$, whose effect in the model is challenging to disentangle from the coefficients \aa\ (but which does not affect the recovery of the linear point).

\subsection{Laguerre model for 2pcf multipoles}
\label{subsec:laguerremodel}
In practice, it is convenient to choose \sfid\ and the range $\{s_{\rm min},s_{\rm max}\}$ over which to assume \eqn{eq:xilin-poly}, based on the observed location and width of the BAO feature in the monopole data, so that $\sfid\simeq10\Mpch$ and $\{s_{\rm min},s_{\rm max}\}\simeq\{60,120\}\Mpch$. While the value of \sfid\ mainly affects numerical stability and does not change any of the results, the final constraints on the parameters \aa\ are mildly sensitive to the choice of $\{s_{\rm min},s_{\rm max}\}$ \citepalias[see][for a discussion]{ps22}.

The expressions for $\xiell{\ell}(s)$ with $\ell=2,4$ depend on volume integrals of $\xi_0$ from $0$ to $s$, while the Laguerre expansion for $\xi_0$ is only valid in the range $s_{\rm min}\leq s\leq s_{\rm max}$. To account for this, we introduce three new free parameters \Qbar{2}{2}, \Qbar{4}{2} and \Qbar{4}{4}, defined by
\be
\Qbar{\ell}{j} \equiv b^2\,\frac{(j+1)}{s_{\rm min}^{j+1}}\int_0^{s_{\rm min}}\der u\,u^j\,\xi_0(u|\sigeff{\ell}{})\,,
\label{eq:Qbar-def}
\ee
so that
\begin{align}
\Qbar{2}{2} &= b^2\,\bar{\xi}_0(s_{\rm min}|\sigeff{2}{})\,,\notag\\
\Qbar{4}{2} &= b^2\,\bar{\xi}_0(s_{\rm min}|\sigeff{4}{})\,,\notag\\
\Qbar{4}{4} &= b^2\,\bar{\bar{\xi}}_0(s_{\rm min}|\sigeff{4}{})\,,
\label{eq:Qbar-interpret}
\end{align}
and define the volume averages of the $\mu_m(x)$ as 
\begin{align}
\bar{\mu}_m(s;\sigma) &\equiv \frac{3}{s^3}\int_{s_{\rm min}}^s\der u\,u^2\,\mu_m(u/\sigma)\,,
\label{eq:mubar-def}\\
\bar{\bar{\mu}}_m(s;\sigma) &\equiv \frac{5}{s^5}\int_{s_{\rm min}}^s\der u\,u^4\,\mu_m(u/\sigma)\,.
\label{eq:mubarbar-def}
\end{align}
Since the lower integration limits in these expressions are $s_{\rm min}$ rather than $0$, we clearly have $\bar{\mu}_m(s_{\rm min};\sigma)=0=\bar{\bar{\mu}}_m(s_{\rm min};\sigma)$. 
Using \eqns{eq:xi0-Lag}, \eqref{eq:mubar-def} and~\eqref{eq:mubarbar-def} to construct $\bar{\xi}_0$ and $\bar{\bar{\xi}}_0$, we can model $\xiell{\ell}(s)$ (equations~\ref{eq:xi0NL-final}-\ref{eq:xi4NL-final}) in the range $s_{\rm min}\leq s\leq s_{\rm max}$ as
\begin{align}
\xiell{0}(s) &= \chi_0\,\sum_{m=0}^{M-1}\frac{a_m}{m!}\left(\frac{\sigeff{0}{}}{\sfid}\right)^m\mu_m(s/\sigeff{0}{})\,,
\label{eq:xiNL(0)-Lag}\\
\xiell{2}(s) &= -\chi_2\bigg[\Qbar{2}{2}\left(\frac{s_{\rm min}}{s}\right)^3 \notag\\
&\ph{b}
- \sum_{m=0}^{M-1}\frac{a_m}{m!}\left(\frac{\sigeff{2}{}}{\sfid}\right)^m\bigg\{\mu_m(s/\sigeff{2}{})-\bar{\mu}_m(s;\sigeff{2}{})\bigg\} \bigg]\,,
\label{eq:xiNL(2)-Lag}\\
\xiell{4}(s) &= \chi_4\bigg[\frac52\,\Qbar{4}{2}\left(\frac{s_{\rm min}}{s}\right)^3 - \frac72\,\Qbar{4}{4}\left(\frac{s_{\rm min}}{s}\right)^5 \notag\\
&\ph{-b^2}
+ \sum_{m=0}^{M-1}\frac{a_m}{m!}\left(\frac{\sigeff{4}{}}{\sfid}\right)^m\bigg\{\mu_m(s/\sigeff{4}{})\notag\\
&\ph{+\sum a_m()^4}
+\frac52\,\bar{\mu}_m(s;\sigeff{4}{})-\frac72\,\bar{\bar{\mu}}_m(s;\sigeff{4}{})\bigg\} \bigg]\,,
\label{eq:xiNL(4)-Lag}
\end{align}
where the prefactors only involve $\chi_\ell$ since we absorbed $b^2$ into the coefficients \aa\ and the definition of \Qbar{\ell}{j}.

To decrease the number of free parameters, we exploit the fact that \Qbar{2}{2} and \Qbar{4}{2} only appear with the combination $(s_{\rm min}/s)^3$, and work with the quantities $\Delta\xiell{\ell}(s)$ defined by
\be
\Delta\xiell{\ell}(s) \equiv \xiell{\ell}(s) - g_\ell\,(s_{\rm min}/s)^3\,\xiell{\ell}(s_{\rm min})\,,
\label{eq:Delta_xiell-def}
\ee
where $g_0 = 0$ and $g_2 = 1 = g_4$.
The models for these quantities are
\begin{align}
\Delta\xiell{0}(s) &= \chi_{0}\bigg[\sum_{m=0}^{M-1}\frac{a_m}{m!}\left(\frac{\sigeff{0}{}}{\sfid}\right)^m\mu_m(s/\sigeff{0}{})\bigg]\,,
\label{eq:Delta_xi0}\\
\Delta\xiell{2}(s) &= \chi_{2}\bigg[\sum_{m=0}^{M-1}\frac{a_m}{m!}\left(\frac{\sigeff{2}{}}{\sfid}\right)^m\bigg\{\mu_m(s/\sigeff{2}{}) \notag\\
&\ph{\chi_2\,\sum_{m=0}^{M-1}}
- (s_{\rm min}/s)^3\,\mu_m(s_{\rm min}/\sigeff{2}{}) -\bar{\mu}_m(s;\sigeff{2}{})\bigg\}\bigg]\,, 
\label{eq:Delta_xi2}\\
\Delta\xiell{4}(s) &= \chi_{4}\bigg[
\frac72\Qbar{4}{4}\left(\frac{s_{\rm min}}{s}\right)^3\left(1-\left(\frac{s_{\rm min}}{s}\right)^2\right) \notag\\
&\ph{\frac72\qbar{4}{4}}
+\sum_{m=0}^{M-1}\frac{a_m}{m!}\left(\frac{\sigeff{4}{}}{\sfid}\right)^m\bigg\{\mu_m(s/\sigeff{4}{}) 
\notag\\ 
&\ph{\frac72\qbar{4}{4}\frac{a_m}{m!}}
- (s_{\rm min}/s)^3\,\mu_m(s_{\rm min}/\sigeff{4}{})  \notag\\ 
&\ph{\frac72\qbar{4}{4}\frac{a_m}{m!}}
 + \frac52\bar{\mu}_m(s;\sigeff{4}{}) - \frac72\bar{\bar{\mu}}_m(s;\sigeff{4}{})\bigg\}\bigg]\,, 
\label{eq:Delta_xi4}
\end{align}
which eliminates the parameters \Qbar{2}{2} and \Qbar{4}{2}. The final dimensionality of our model is then $M+3$ when only using the monopole and quadrupole and $M+4$ when using all $\ell=0,2,4$. Since $\Delta\xiell{\ell}(s_{\rm min})=0$ by construction, we delete the corresponding two data points for $\ell=2$ and $4$ from the data set after constructing $\Delta\xiell{\ell}(s)$ from the observed $\xiell{\ell}(s)$. 

Written like this, the model explicitly depends on the parameter $b$ only through the definitions of \sigeff{\ell}{2} which contain factors of $f=b\,\beta$. Since this dependence is quite weak, we include in our data set a measurement of the integral $\Sigma_{\rm obs}^2$ of the monopole power spectrum in linearly spaced bins of \kk,
\be
\hat\Sigma_{\rm obs}^2 \equiv \frac{\Delta k}{6\pi^2}\sum_i\,P^{(0)}(k_i)\,,
\label{eq:Sigobs2}
\ee
which we model using
\begin{align}
\Sigma_{\rm obs}^2 &= \frac13\int\der\ln k\, k^{-2}\,\Dellsq{0}(k)\notag\\
&\approx \frac13\,\beff{0}{2}\int\der\ln k\, k^{-2}\,\Delta_{\rm lin}^2(k)\,\e{-k^2\sigeff{0}{2}/2}\notag\\
&\approx  b^2\,\chi_0(\beta)\,\sigv^2\,.
\label{eq:Sigobs2-model}
\end{align}
For the BOSS Final Year cosmology, the model approximation in the last line gives $\Sigma_{\rm obs}=10.9 \,(10.0)\,\Mpch$, compared to $9.3 \,(8.4)\,\Mpch$ from the exact integral in the first line for DESI (BOSS DR12) LRGs, an overestimate by $\sim17\%\,(19\%)$. We therefore divide the last line of \eqref{eq:Sigobs2-model} by $1.38$ to account for the overestimate, 
\be
\Sigma_{\rm obs,model}^2 = b^2\,\chi_0(\beta)\,\sigv^2\,/\,1.38\,.
\label{eq:Sigobs2-model-hack}
\ee
We discuss this limitation of our model in section~\ref{subsec:caveats} below.

\subsection{Gauss-Poisson covariance matrix}
For simplicity, in this work we use the Gauss-Poisson approximation to model the covariance matrix of our mock observations. This can be easily replaced with more accurate covariance matrix estimates based on simulations or mock galaxy catalogs, as needed \citep[e.g.,][]{chuang+15,zhao+21}.

In the Gauss-Poisson approximation, the covariance matrix
\be
C^{\ell\ell^\prime}_{ij} \equiv \avg{\xiell{\ell}(s_i)\xiell{\ell^\prime}(s_j)} - \avg{\xiell{\ell}(s_i)}\avg{\xiell{\ell^\prime}(s_j)}
\label{eq:covmat-def}
\ee
can be written as \citep[e.g.][]{grieb2016},
\begin{equation}
C^{\ell\ell^\prime}_{ij} = \frac{i^{\ell_1 + \ell_2}}{2\pi^2}\int dk\,k^2\,{\bar j}_{\ell_1}(ks_i)\,{\bar j}_{\ell_2}(ks_j) \,\sigma^2_{\ell_1\ell_2}(k)\,,
\label{eq:C_ell1ell2(si,sj)}
\end{equation}
 where
\begin{align}
\sigma^2_{\ell_1\ell_2}(k) &= \frac{(2\ell_1 + 1)(2\ell_2 + 1)}{V_{\rm sur}/2}\notag\\
&\ph{V_{sur}/2}
\times\int_{-1}^{1} \frac{d\mu}{2}\,
\left[P(k,\mu) + \frac{1}{\bar{n}}\right]^2\,\Pell{\ell_1}(\mu)\,\Pell{\ell_2}(\mu) 
\label{eq:sig2_ell1ell2(k)}
\end{align}
for a survey of volume $V_{\rm sur}$ with observed tracer number density $\bar n$, and 
\be
{\bar j}_{\ell}(ks_i) = \frac{4\pi}{V_i}\int ds\,s^2\,j_\ell(ks)\,W_i(s)
\label{eq:jellbar}
\ee
with
\be
V_i = 4\pi \int ds\,s^2\,W_i(s)\,,
\ee
where $W_i(s)$ describes the shape of a window over which $j_\ell$ has been averaged.  E.g., for the tophat bins of width $\Delta s$ centered on $s_i$ that we use, $W_i(s) = 1$ if $s_i-\Delta s/2 \le s\le s_i + \Delta s/2$, and the integral which defines $\bar{j}_\ell$ can be done analytically.

Since we use $\Delta\xiell{\ell}$ (equations~\ref{eq:Delta_xi0}-\ref{eq:Delta_xi4}) as our observables instead of \xiell{\ell}, the covariance matrix $C^{\ell\ell^\prime}_{ij}$ used in the likelihood evaluation must be replaced with
\begin{align}
\Delta C^{\ell\ell^\prime}_{ij} &= C^{\ell\ell^\prime}_{ij} - g_\ell\left(\frac{s_1}{s_i}\right)^3C^{\ell\ell^\prime}_{1j} - g_{\ell^\prime}\left(\frac{s_1}{s_j}\right)^3C^{\ell\ell^\prime}_{i1} \notag\\
&\ph{ C^{\ell\ell^\prime}_{ij}- g_{\ell^\prime}}
+ g_\ell\, g_{\ell^\prime}\left(\frac{s_1^2}{s_is_j}\right)^3C^{\ell\ell^\prime}_{11}\,,
\label{eq:Delta_Cellellpr}
\end{align}
where we assumed $s_1=s_{\rm min}$ and it is understood that we discard the rows and columns corresponding to $s=s_{\rm min}$ for $\ell=2$ and $\ell=4$.

Finally, when including in the data set a measurement $\Sigma_{\rm obs}^2$ (equation~\ref{eq:Sigobs2}),
we must modify the covariance matrix further, by including a row and column accounting for the error in $\Sigma_{\rm obs}^2$ and its covariance with the measured $\xiell{\ell}(s)$ in bins of $s$. In the Gauss-Poisson approximation, these are respectively given by
\begin{align}
&{\rm Var}\left(\hat\Sigma_{\rm obs}^2\right) = \frac{8\Delta k^2}{9}\sum_i\frac{1}{V_{k_i}^2}\int_{k_i-\Delta k/2}^{k_i+\Delta k/2}\der k\,k^2\sigma^2_{00}(k)\,,
\label{eq:Var(Sig2obs)}\\
&{\rm Cov}\left(\hat\Sigma_{\rm obs}^2,\xiell{\ell}(s_j)\right) \notag\\ 
&= \frac{2\Delta k}{3\pi}\,i^\ell\sum_i\frac{1}{V_{k_i}}\int_{k_i-\Delta k/2}^{k_i+\Delta k/2}\der k\,k^2\,\bar{j}_\ell(ks_j)\,\sigma^2_{0\ell}(k)\,,
\label{eq:Cov(Sig2obs,xiell)}
\end{align}
where $V_{k_i} = (4\pi/3)[(k_i+\Delta k/2)^3 - (k_i-\Delta k/2)^3]$ and $\sigma^2_{\ell_1\ell_2}(k)$ and $\bar{j}_\ell(ks_j)$ are defined in \eqns{eq:sig2_ell1ell2(k)} and \eqref{eq:jellbar}, respectively. To calculate the covariance of $\hat\Sigma_{\rm obs}^2$ with $\Delta\xiell{\ell}$ instead of $\xiell{\ell}$, we simply subtract $g_\ell(s_1/s_j)^3{\rm Cov}(\hat\Sigma_{\rm obs}^2,\xiell{\ell}(s_1))$ from \eqn{eq:Cov(Sig2obs,xiell)} and delete the entries corresponding to $s_1=s_{\rm min}$ as before.

\section{Analysis}
\label{sec:analysis}
We use mock data generated assuming two choices of sample: (i) a BOSS DR12 LRG-like sample and (ii) a DESI LRG-like sample. In each case, we define a multivariate Gaussian using \eqn{eq:xiell-2} and the first line of \eqn{eq:Sigobs2-model} as the mean, and \eqns{eq:C_ell1ell2(si,sj)}, \eqref{eq:Var(Sig2obs)} and~\eqref{eq:Cov(Sig2obs,xiell)} with tophat binning as the covariance matrix. The corresponding choices of the survey redshift $z_{\rm sur}$, effective survey volume $V_{\rm sur}$, galaxy number density $\bar{n}$ and linear halo bias $b$ are summarised in Table~\ref{tab:toy_samples}. For each survey choice, we draw one realisation from the corresponding multivariate Gaussian and construct the observables $\Delta\xiell{\ell}(s)$ using \eqn{eq:Delta_xiell-def} for $55\Mpch\leq s\leq 125\Mpch$ in $72$ linearly spaced, $\simeq1\Mpch$ wide bins, along with $\hat\Sigma_{\rm obs}^2$ in linearly spaced bins $k_{\rm min} < k_i < 0.5\,h\,{\rm Mpc}^{-1}$ with spacing $\Delta k = k_{\rm min}/2$, where $k_{\rm min}=2\times2\pi/V_{\rm sur}^{1/3}$. This gives 84 (189) $k$-bins for the BOSS DR12 (DESI) configuration. 

The BOSS DR12 numbers in Table~\ref{tab:toy_samples} are chosen to match the `Bin 3' sample from \citet{BOSSDR12-FinalData}, while the DESI LRG numbers are consistent with \citet[][see their section 3.2]{DESI} and \citet{zhou+20}. The linear binning choice of $k$-bins for the covariance of $\hat\Sigma_{\rm obs}^2$ is similar to that in fig.~3 of \citet{beutler+14}. All the results use the BOSS Final Year flat $\Lambda$CDM cosmology \citep{alam+17} with parameters as listed in the Introduction. The input (`true') values of $(f,\sigv)$ for the BOSS DR12 and DESI configurations are, respectively, $(0.790,4.19\,\Mpch)$ and $(0.814,4.01\,\Mpch)$.

\begin{table}
\centering
\begin{tabular}{ccccc}
\hline\hline
mock & $z_{\rm sur}$ & $V_{\rm sur}$ & $\bar{n}$  & $b$ \\
sample &  & (${\rm Gpc}^3$) & $10^{-4}\,(\Mpch)^{-3}$ &\\
\hline 
BOSS DR12 & $0.61$ & $4.1$ & $4.7$  & $2.1$ \\
DESI LRG & $0.7$ & $45.3$ & $6.0$ & $2.435$ \\
\hline\hline
\end{tabular}
\caption{Sample definitions for mock measurements. The columns give the assumed values of effective survey redshift $z_{\rm sur}$, survey volume $V_{\rm sur}$ in ${\rm Gpc}^3$, galaxy number density in $10^{-4}\,(\Mpch)^{-3}$ and linear bias $b$, for the mock samples representing BOSS DR12 LRGs and DESI LRGs. See text for original references.}
\label{tab:toy_samples}
\end{table}

\subsection{Bayesian sampling}
In this section, we describe our setup for Bayesian sampling using the Markov Chain Monte Carlo (MCMC) technique to constrain the model parameters using the mock data described above.

\subsubsection{Likelihood}
We construct a Gaussian likelihood of the form ${\rm e}^{-\chi^2/2}$, with 
\be
\chi^2 = (\mathbf{d} - \mathbf{m})^{\rm T}\, C^{-1}\, (\mathbf{d} - \mathbf{m})\,,
\label{eq:chi2}
\ee
where $\mathbf{d}$ and $\mathbf{m}$ represent the data and model vector, respectively, and $C$ is the data covariance matrix. 

The data vector $\mathbf{d}$ comprises the $\Delta\xiell{\ell}$ measurements and $\hat\Sigma_{\rm obs}^2$ as described above. For the BOSS DR12 configuration, we use $\ell=0$ and $2$ measurements for $\Delta\xiell{\ell}$, since $\ell=4$ is not measured with enough precision to yield useful information. For the DESI configuration, on the other hand, we use $\ell=0,2$ and $4$. Keeping in mind the deletion of data points when constructing $\Delta\xiell{\ell}$ from \xiell{\ell} (see the discussion below equation~\ref{eq:Delta_xi4}), this leads to data vectors of length $144$ and $215$ for the BOSS DR12 and DESI configurations, respectively. The data covariance $C$ is correspondingly modelled by combining \eqns{eq:Delta_Cellellpr}, \eqref{eq:Var(Sig2obs)} and \eqref{eq:Cov(Sig2obs,xiell)}, along with the modification  discussed below \eqn{eq:Cov(Sig2obs,xiell)}.

The model vector $\mathbf{m}$ correspondingly comprises \eqns{eq:Delta_xi0}-\eqref{eq:Delta_xi4} for $\Delta\xiell{\ell}$ and \eqn{eq:Sigobs2-model-hack} for $\Sigma_{\rm obs}^2$. For reasons we discuss below, we use a degree 7 polynomial ($M=8$) for the BOSS DR12 configuration to model the linear theory 2pcf in \eqref{eq:xilin-poly}, while for the DESI configuration we use a degree 9 polynomial ($M=10$). Along with the three parameters $b$, $\beta$ and \sigv, and the new parameter \Qbar{4}{4} when using $\ell=4$ measurements, this leads to $133$ ($201$) degrees of freedom for the BOSS DR12 (DESI) configuration.

\subsubsection{Eigen-coefficients and standardisation}
Unlike \citetalias{ps22}, we have chosen not to introduce an offset scale $r_{\rm fid}$ in \eqn{eq:xilin-poly} -- which would replace $r^m\to(r-r_{\rm fid})^m$ -- since this considerably simplifies the manipulations of the Laguerre functions $\mu_m(x)$. This, however, introduces strong degeneracies between the polynomial coefficients \aa, because the 2pcf data lie very far from $r=0$. These degeneracies, if not handled carefully, render the MCMC sampling unstable.

To deal with this, we approximately decorrelate the polynomial coefficients prior to MCMC sampling, as follows. We first fix the parameters $(b,\beta,\sigv)$ to some fiducial values $(b_\ast,\beta_\ast,\sigma_{{\rm v}\ast})$ and \emph{analytically} solve the linear Gaussian problem for the \emph{a posteriori} distribution of \aa\ using the monopole data \xiell{0} alone, exactly as described by \citetalias{ps22}. Since this distribution is a multivariate Gaussian in \aa, at this stage we obtain a mean vector $\aa_\ast$ and an $M$-dimensional covariance matrix $C^{\rm (poly)}$ for the polynomial coefficients \aa. The diagonalising rotation $R$ of $C^{\rm (poly)}$ then approximately decorrelates \aa\ by defining
\be
\eeb = R^{\rm T}\cdot \aa
\label{eq:eigenvec}
\ee
Hereon, we refer to \eeb\ as the \emph{eigen-coefficients} corresponding to \aa\ even though, strictly speaking, this decorrelation is only approximate when we also vary $b,\beta,\sigv$ below. We also emphasize that, although the rotation $R$ is defined for a specific choice of fiducial values $(b_\ast,\beta_\ast,\sigma_{{\rm v}\ast})$, this is a fixed rotation introduced purely for convenience. Our approach does not rely on these fiducial values being close to the `truth'; values very different from the underlying true model would only result in somewhat longer convergence times for the MCMC chains, but will not bias the final result.

Another issue we must deal with is that the typical values of the best fit eigen-coefficients span a large dynamic range of $\gtrsim10^8$. To simplify the MCMC sampling, we therefore define the standardised eigen-coefficients $\tilde{\eeb}$ using
\be
\tilde{e}_m = (e_m - \avg{e}_m)/\sigma_{m} \,;\quad 0\leq m \leq M-1\,,
\label{eq:eigvec-std}
\ee
where $\avg{e}_m$ and $\sigma_m$ are the mean and standard deviation, respectively, of the $m^{\rm th}$ eigen-coefficient as obtained from the linear Gaussian analysis. Similarly, for convenience, we also standardise the parameters $(b,\beta,\sigv)$ by defining
\begin{align}
\tilde{b} = b/b_\ast - 1\,; \quad \tilde{\beta} = \beta/\beta_\ast - 1\,; \quad \tilde{\sigma}_{\rm v} = \sigv/\sigma_{{\rm v}\ast}-1\,,
\end{align}
although this is not as essential as the standardisation of \eeb, since the parameters $(b,\beta,\sigv)$ are typically of order unity. When including $\ell=4$ data, we similarly standardise \Qbar{4}{4} using
\be
\tilde{Q}^{(4)}_4 = (\Qbar{4}{4} - Q_\ast)/\sigma_{Q\ast}\,,
\label{eq:Q-std}
\ee
where $Q_\ast$ and $\sigma_{Q\ast}$ are respectively set equal to the expected typical value (we use $Q_\ast=0.008\times b_\ast^2$ throughout) and width of the prior range described later.

We set up the MCMC described below to sample the parameters $\{\tilde{\eeb},\tilde{b},\tilde{\beta},\tilde{\sigma}_{\rm v},[\tilde{Q}^{(4)}_4]\}$, but display all our final results in terms of $\{\eeb,b,\beta,\sigv,[\Qbar{4}{4}]\}$ by simply undoing the standardisations. 

\subsubsection{Parameter priors}
We define priors on the sampled parameters as follows. 

For each eigen-coefficient $e_m$ with $0\leq m\leq M-1$, we use broad uniform priors in the range $e_m\in[\avg{e_m}-25\sigma_m,\avg{e_m}+25\sigma_m]$, where $\avg{e_m}$ and $\sigma_m$ are obtained from the linear Gaussian analysis as described above. For the parameters $\beta$ and \sigv, we use broad uniform priors over the ranges $\beta\in[-10,+10]$ and $\sigv\in[0,+20]\,\Mpch$. When including $\ell=4$ data, we use a uniform prior on \Qbar{4}{4} in the range $\Qbar{4}{4}\in [0,+0.1]\times b_\ast^2$ (so that $\sigma_{Q\ast}=0.1\,b_\ast^2$ in equation~\ref{eq:Q-std}).

For the linear bias $b$, we will display results assuming a $10\%$ Gaussian prior, namely, a Gaussian distribution with mean $b_{\rm true}$ and standard deviation $0.1\,b_{\rm true}$, where $b_{\rm true}$ is the appropriate input value from Table~\ref{tab:toy_samples}. We discuss this choice further in section~\ref{subsec:caveats}, and later also report results of relaxing the prior to be uniform in the range $b\in[-20,+20]$.

\subsubsection{MCMC sampling}
We use the publicly available Python-based framework \textsc{cobaya} \citep{tl19-cobaya,tl21-cobaya}\footnote{\url{https://cobaya.readthedocs.io/}} to perform MCMC sampling. We use the \texttt{Theory} class of \textsc{cobaya} to set up numerical evaluations of \eqns{eq:Delta_xi0}-\eqref{eq:Delta_xi4} and \eqn{eq:Sigobs2-model-hack} for the model, and the \texttt{Likelihood} class to evaluate \eqref{eq:chi2} for the the log-likelihood. Throughout, we sample the standardised parameters $\{\tilde{\eeb},\tilde{b},\tilde{\beta},\tilde{\sigma}_{\rm v},[\tilde{Q}^{(4)}_4]\}$, appropriately accounting for the standardisation in the model evaluation. We set the fiducial parameter values $(b_\ast,\beta_\ast,\sigma_{\rm v\ast})$ to the input values for each mock sample, namely, $(2.1,0.376,4.19\Mpch)$ for BOSS DR12 LRGs and $(2.435,0.334,4.01\Mpch)$ for DESI LRGs.  

\begin{figure*}
\centering
\includegraphics[width=0.45\textwidth,trim=5 10 5 5,clip]{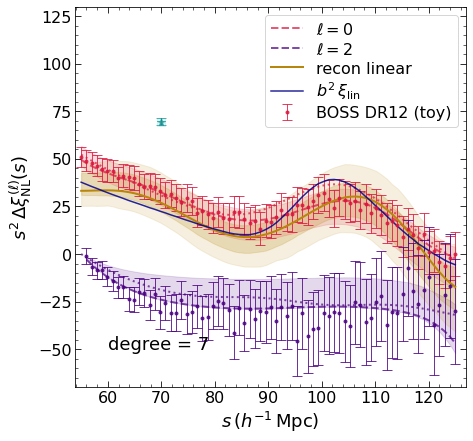}
\includegraphics[width=0.435\textwidth,trim=5 10 5 5,clip]{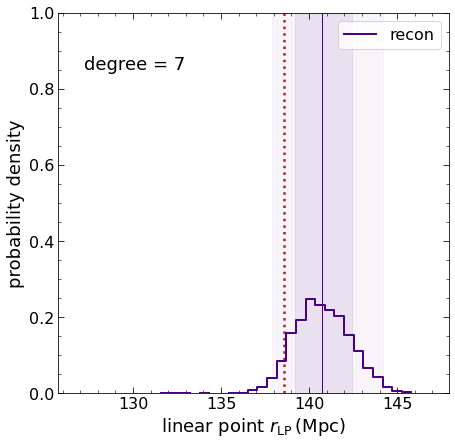}
\caption{
\emph{(Left panel):} Comparison of mock data and best fit models for $s^2\Delta\xiell{\ell}(s)$ with $\ell=0,2$, along with $\Sigma_{\rm obs}^2$, for the mock BOSS DR12 data (Table~\ref{tab:toy_samples}) analysed with a degree 7 Laguerre function (section~\ref{subsec:laguerremodel}). Data points with errors show the mock measurements, dashed curves show the best fit model, with the shaded bands indicating the respective central $68\%$ confidence region. 
The single cyan data point with error bar shows the mock value of $\Sigma_{\rm obs}^2$, with the shaded cyan box (nearly indistinguishable from the data point), showing the value in the best fit model.
Dotted curves show the input `truth' used to generate the data as described in section~\ref{sec:analysis}. Yellow solid curve shows the reconstructed linear theory $\xi_{\rm L}$ using the best fit polynomial coefficients, with the inner (outer) yellow bands showing the central $68\%$ ($95\%$) confidence region. Blue solid curve shows the linear theory $\xi_{\rm L}=b^2\xi_{\rm lin}$ in the input cosmology. 
\emph{(Right panel):} Recovery of the linear point $r_{\rm LP}$ for the same analysis. Histogram shows the inferred distribution of $r_{\rm LP}$, with the solid vertical line and inner (outer) vertical band showing the median and central $68\%$ ($95\%$) confidence region. Dotted red vertical line shows the value in the input cosmology. This figure can be compared with figs.~5 and~6 of \citetalias{ps22}, where the analysis only used $\ell=0$ and did not vary the parameters $\{b,\beta,\sigv\}$.}
\label{fig:mock_bossdr12_xisLP}
\end{figure*}

During the model evaluation, we  internally scale all the data by a constant factor $10^4$ in order to numerically stabilise the initial linear Gaussian calculation \citepalias[see][for a discussion]{ps22}, and undo the scaling when presenting results. The likelihood calculation also enforces $f = b\,\beta \geq 0$, by returning zeros for all $\Delta\xiell{\ell}(s)$ otherwise. We use the \texttt{mcmc} sampler included in \textsc{cobaya}, which implements the Metropolis-Hastings algorithm. We initialise all chains by sampling narrow uniform distributions in each parameter, centered on the linear Gaussian result, and stop the chains when the Gelman-Rubin index $|R-1|$ falls below 0.05 for convergence of both, means as well as $95\%$ bounds. The MCMC chains are analysed using the Python package \textsc{getdist} \citep{lewis19},\footnote{\url{https://getdist.readthedocs.io/}} discarding the first $30\%$ of the samples as burn-in.

\section{Results}
\label{sec:results}
We now discuss the results of the MCMC analysis of the BOSS DR12 and DESI LRG mock data.

\subsection{Parameter constraints}
\label{subsec:constraints}
Fig.~\ref{fig:mock_bossdr12_xisLP} shows the best fit $\Delta\xiell{\ell}(s)$ and and $\Sigma_{\rm obs}^2$ \emph{(left panel)} and the inferred distribution of the linear point $r_{\rm LP}$ \emph{(right panel)}, by fitting a degree $7$ Laguerre function to the mock BOSS DR12 measurements for $\ell=0$ and $2$ (shown as the data points with errors). We discuss the choice of degree later.

\begin{figure*}
\centering
\includegraphics[width=0.95\textwidth,trim=5 10 5 5,clip]{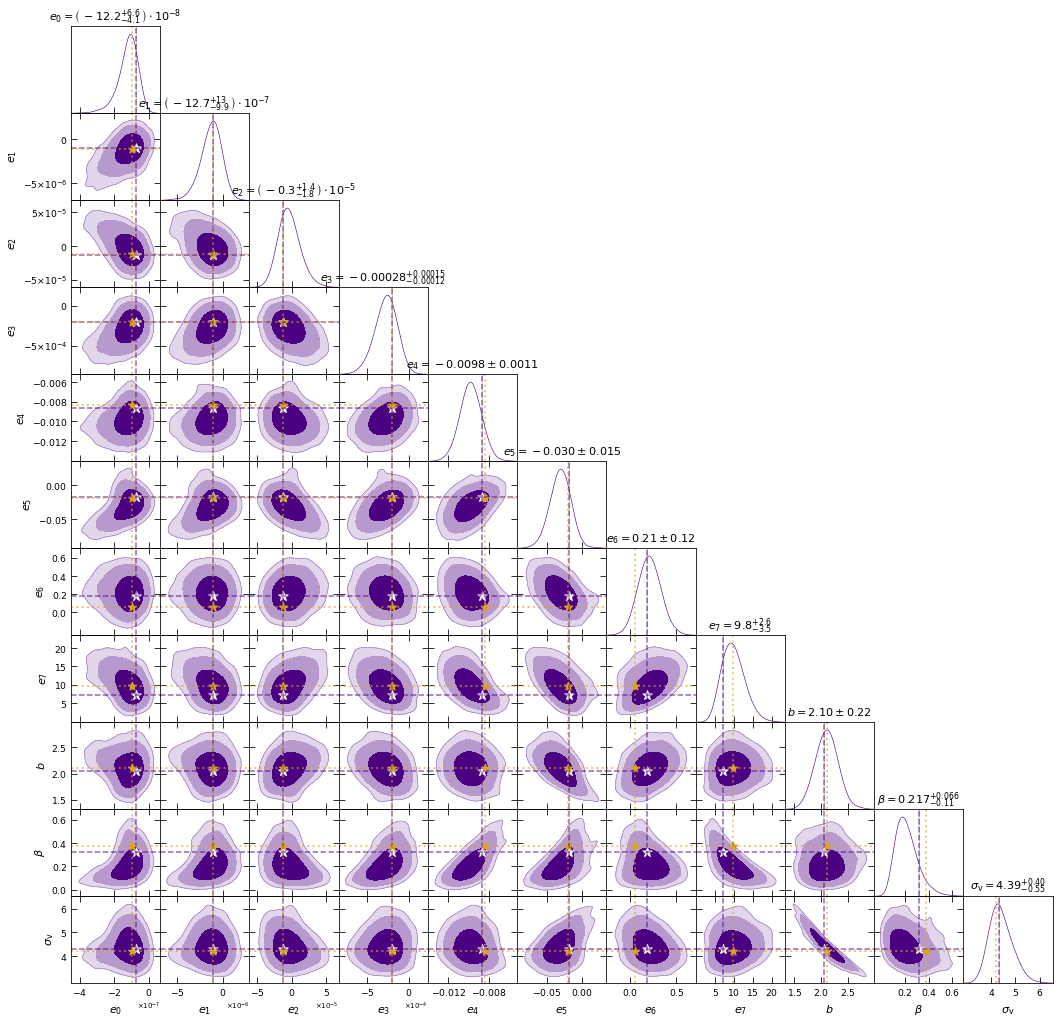}
\caption{Joint \emph{a posteriori} distributions of all (eigen)-parameters varied to obtain the results in Fig.~\ref{fig:mock_bossdr12_xisLP} for the BOSS DR12 mock data set. Dashed purple lines intersecting at white stars show the best fit values. Yellow dotted lines intersecting at yellow stars show the input values for $b$, $\beta$ and $\sigv$ and the linear Gaussian prediction (obtained at the input $b$, $\beta$, \sigv) for the eigen-coefficients $e_p$.}
\label{fig:mock_bossdr12_contours}
\end{figure*}

\begin{table}
\centering
\begin{tabular}{cccccc}
\hline\hline
mock & $b$ & $\beta$ & \sigv\   & \Qbar{4}{4} & $\chi^2/{\rm dof}$ \\
 &  &  & (\Mpch) & & \\
\hline 
BOSS DR12 & $2.04$ & $0.323$ & $4.31$  & - & $128.9/133$ \\
DESI LRG & $2.52$ & $0.234$ & $4.00$ & $0.081$ & $202.9/201$ \\
\hline\hline
\end{tabular}
\caption{Best-fit values of the parameters $\{b,\beta,\sigv,[\Qbar{4}{4}]\}$ and $\chi^2$ per degree of freedom from the MCMC analysis of the BOSS DR12 and DESI LRG mock data. See Fig.~\ref{fig:mock_bossdr12_contours} and Fig.~\ref{fig:mock_desilrg_contours} for the median and central $68\%$ confidence ranges of each parameter in the respective samples.}
\label{tab:bestfit}
\end{table}

Fig.~\ref{fig:mock_bossdr12_contours} shows the corresponding \emph{a posteriori} distributions of the varied parameters, with Table~\ref{tab:bestfit} summarising the best-fit values of $\{b,\beta,\sigv\}$ and the best fit $\chi^2$ per degree of freedom. 
We see a strong degeneracy between $b$ and \sigv, which our model and dataset are unable to break.
Nevertheless, the constraints on 
all parameters, including the derived linear point, comfortably include the input values at better than $95\%$ confidence, so that our method should be unbiased for a BOSS DR12 LRG-like sample. 
While the linear bias is constrained with $10\%$ uncertainty by choice of prior, the $1\sigma$ uncertainties on $\beta$ and \sigv\ are $\sim40\%$ and $\sim10\%$, respectively, relative to the corresponding median. For the linear point $r_{\rm LP}$ (\emph{right panel} of Fig.~\ref{fig:mock_bossdr12_xisLP}), the median and central $68\%$ confidence interval is $r_{\rm LP}=140.7^{+1.7}_{-1.5}$ Mpc, i.e, a $1\%$ uncertainty. We discuss the effect of opening up the prior on $b$ in section~\ref{subsec:caveats}.

One feature of the best fit model, seen already in Fig.~\ref{fig:mock_bossdr12_xisLP}, is that the best fit $\Delta\xiell{2}$ (dashed purple curve) lies at the edge of the $1\sigma$ (purple) band of the prediction in data space. This is unlike $\Delta\xiell{0}$ for which the $1\sigma$ (red) band is symmetrically placed around the best fit (red dashed) curve. This already indicates that our model approximation for $\ell=2$ is approaching the limit of its validity. Another hint that this is happening comes from the fact that the marginal \emph{a posteriori} distribution of $\beta$ in Fig.~\ref{fig:mock_bossdr12_contours} is shifted to substantially lower values as compared to the best fit and input values of $\beta$. We discuss this issue further in section~\ref{subsec:caveats} as well.




\begin{figure*}
\centering
\includegraphics[width=0.45\textwidth,trim=5 10 5 5,clip]{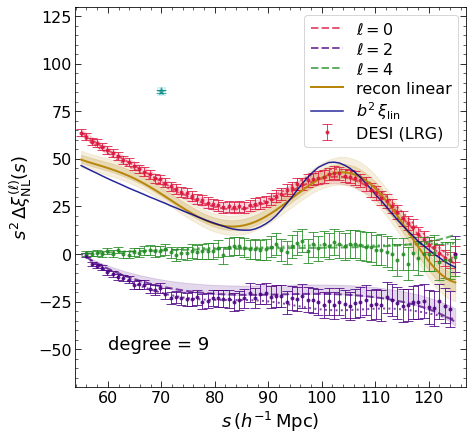}
\includegraphics[width=0.435\textwidth,trim=5 10 5 5,clip]{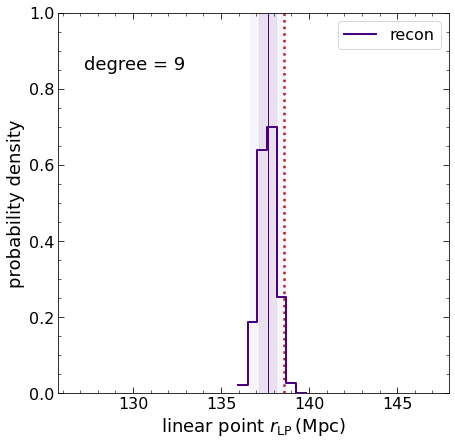}
\caption{Similar to Fig.~\ref{fig:mock_bossdr12_xisLP}, showing results for the mock DESI LRG data (Table~\ref{tab:toy_samples}) for $s^2\Delta\xiell{\ell}(s)$ with $\ell=0,2,4$, along with $\Sigma_{\rm obs}^2$, analysed with a degree 9 Laguerre function. 
}
\label{fig:mock_desilrg_xisLP}
\end{figure*}

The choice of a degree 7 polynomial needs some discussion, since the linear Gaussian analysis of the monopole \xiell{0} by \citetalias{ps22}, using fixed values of $b$, $\beta$ and \sigv, indicated that degree 3 should be sufficient for the BOSS DR12 LRG sample. Our analysis, on the other hand, includes the $\ell=2$ data and allows $\{b,\beta,\sigv\}$ to vary.
We have checked that, using degree 3, 5 and 9 polynomials leads to poorer constraints on $r_{\rm LP}$ ($1\sigma$ errors of $\sim9.5$ Mpc, $\sim2.5$ Mpc and $\sim1.9$ Mpc, respectively) as compared to degree 7 ($\sim1.6$ Mpc, see above). This indicates that the inclusion of $\ell=2$, along with the opening up of the parameter directions $b$, $\beta$ and \sigv, forces us towards higher degree polynomials, with an optimal degree that leads to the minimum error on $r_{\rm LP}$.\footnote{In each of these cases, we found the same behaviour of the best fit $\Delta\xiell{2}$ compared with the central $68\%$ confidence range, as described above, implying a similar level of systematic error in the $\ell=2$ model.}
We have also checked that including $\ell=4$ measurements in the analysis \emph{also} degrades the constraints on $r_{\rm LP}$, by broadening the $68\%$ ($95\%$) confidence region by a factor $\sim2.3\,(1.5)$. This shows that the hexadecapole in BOSS DR12 is expected to add mainly noise rather than signal to the data set \citep[c.f.,][]{rpm15}.



\begin{figure*}
\centering
\includegraphics[width=0.95\textwidth,trim=5 10 5 5,clip]{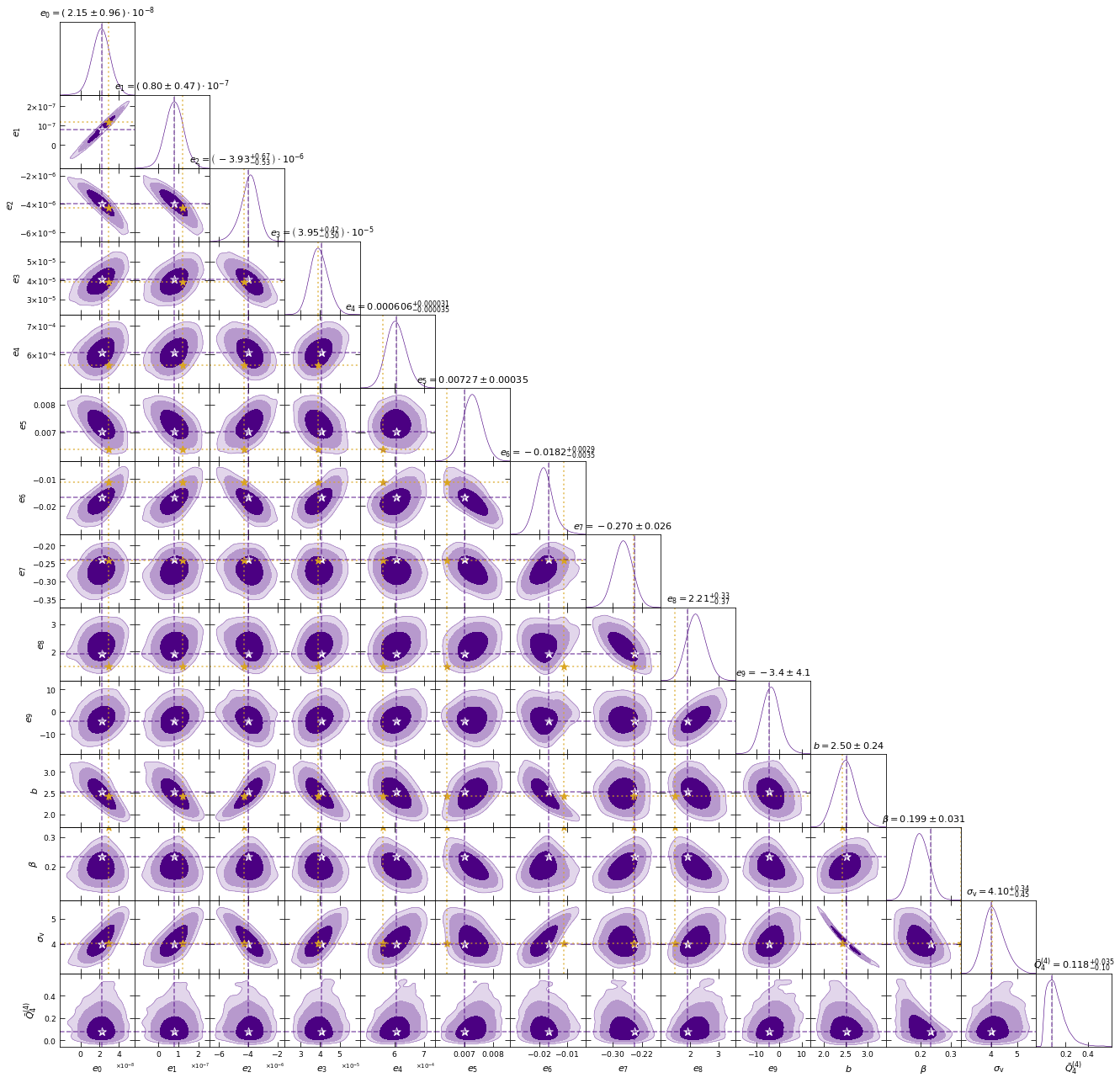}
\caption{Similar to Fig.~\ref{fig:mock_bossdr12_contours}, for the mock DESI LRG data shown in Fig.~\ref{fig:mock_desilrg_xisLP}. In this case, an additional parameter \Qbar{4}{4} was varied in the MCMC analysis (see sections~\ref{subsec:laguerremodel} and~\ref{subsec:constraints}).}
\label{fig:mock_desilrg_contours}
\end{figure*}

Fig.~\ref{fig:mock_desilrg_xisLP} is formatted identically to Fig.~\ref{fig:mock_bossdr12_xisLP} and shows the results of the analysis with the mock DESI LRG measurements of multipoles $\ell=0,2$ and $4$, now with a degree 9 Laguerre function, the choice of degree being made similarly to that described above for the BOSS DR12 mock data. Fig.~\ref{fig:mock_desilrg_contours} shows the corresponding \emph{a posteriori} distributions, with Table~\ref{tab:bestfit} summarising the best-fit values of $\{b,\beta,\sigv,\Qbar{4}{4}\}$ and the best fit $\chi^2$ per degree of freedom.
The degeneracy between $b$ and \sigv\ is now more pronounced, largely driven by the reduced error on $\hat\Sigma_{\rm obs}^2$.
The $1\sigma$ uncertainty on $b$ is again $\sim10\%$ by choice of prior, with the constraint being unbiased relative to the input value. The best-fit and median values of \sigv\ agree well with the input value, and the $\sim10\%$ uncertainty is the same as for the BOSS DR12 mock sample.
The $\sim16\%$ uncertainty on $\beta$ is substantially smaller than for BOSS DR12, but the distribution of $\beta$ excludes the input value at $\gtrsim99\%$ confidence. Alongside, the best fit values of many of the eigen-coefficients \eeb\ are also significantly far from the linear Gaussian expectation at fixed $b$, $\beta$ and \sigv, although each is well-constrained. This is connected to the behaviour of the $\beta$ constraints we highlighted above for the BOSS DR12 mock analysis, and we discuss it further in section~\ref{subsec:caveats}.
The auxiliary parameter \Qbar{4}{4} (which has a non-negative prior) is reasonably well-constrained, but with a long positive tail. 
Finally, the linear point $r_{\rm LP}$ (\emph{right panel} of Fig.~\ref{fig:mock_desilrg_xisLP}) is recovered with very high precision and reasonable accuracy: the median and central $68\%$ confidence interval are $r_{\rm LP}=137.7^{+0.5}_{-0.5}$ Mpc, i.e, a $0.4\%$ uncertainty, with the input value being inside the $95\%$ confidence region. 

\subsection{Theoretical uncertainties and possible improvements}
\label{subsec:caveats}
In this section, we discuss some of the limitations of the approximations made in our model and briefly indicate some directions for improvement.

\subsubsection{Prior on linear bias}
Our results above used a $10\%$ Gaussian prior on the linear bias $b$, which requires some discussion. As mentioned previously, $b$ appears explicitly in our parametrisation only in the evaluation of \sigeff{\ell}{} (equations~\ref{eq:sigeff-def}), whose dependence on $b$ through explicit factors of $f$ is quite weak, and in the model for $\Sigma_{\rm obs}^2$ (equation~\ref{eq:Sigobs2-model-hack}), where $b$ is highly degenerate with \sigv\ and also, to some extent, with $\beta$. (The degeneracy with \sigv\ is closely connected to the well-known degeneracy between $b$ and $\sigma_8$ in usual cosmological analyses of 2pcf observations.) 

It is therefore important to ask how our results would be affected if we did not include a strong prior on $b$.
For the BOSS DR12 configuration, we have checked that opening up the prior on $b$ to the range $b\in[-20,+20]$ (instead of a $10\%$ Gaussian prior), leads to overall weaker constraints on all parameters which are still unbiased at $95\%$ confidence relative to the input values, with the uncertainties on $b$, $\beta$ and \sigv\ being, respectively, $\sim37\%$, $\sim40\%$ and $\sim42\%$. The uncertainties on the eigen-coefficients are typically a factor $2$ larger than those seen in Fig.~\ref{fig:mock_bossdr12_contours}. The corresponding uncertainty on $r_{\rm LP}$, however, is again $\sim1\%$. 

In other words, the main effect of opening up the parameter space along $b$ is to degrade the constraint on $b$ itself and on \sigv\ (as expected from their degeneracy), but not on $\beta$ or the linear point. 
Having said this, we also expect that the combination of galaxy positions and weak gravitational lensing from upcoming surveys should be able to constrain the linear bias to an accuracy of $\sim10\%$, by breaking the degeneracy between $b$ and $\sigma_8$ in those observations \citep{miyatake+22,pandey+22}. In this case, a $10\%$ Gaussian prior on $b$ would not be unrealistic.

We therefore conclude that our choice of prior on $b$ is not likely to be a cause for concern in our framework.

\subsubsection{Constraint on $\beta$}
As we discussed in section~\ref{subsec:modelaccuracy}, we expect our model to become increasingly inaccurate with increasing $\ell$, for sufficiently precise observations (\emph{right panel} of Fig.~\ref{fig:diagnostic}).
This is particularly clear from the behaviour of the constraint on $\beta$ in the DESI LRG configuration presented in section~\ref{subsec:constraints}.
For each $\ell$, $\beta$ appears in the model for $\Delta\xiell{\ell}$ primarily through a multiplicative factor of $\chi_\ell(\beta)$  (equations~\ref{eq:Delta_xi0}-\ref{eq:Delta_xi4}). Since $\chi_0=1+\Cal{O}(\beta)$, $\chi_2=\Cal{O}(\beta)$ and $\chi_4=\Cal{O}(\beta^2)$, it is immediately obvious that $\beta$ constraints using high-precision measurements of $\ell=2$ and especially $\ell=4$ will be increasingly sensitive to systematic errors in the model. For example, we have checked that the relative observational errors expected for the DESI LRG $\xiell{4}(s)$ are \emph{smaller} than the relative systematic errors in the model (c.f., Fig.~\ref{fig:diagnostic}) by a factor $\gtrsim2$ for $s\lesssim90\Mpch$. Consistently with this, we find that \emph{excluding} the $\ell=4$ data from the DESI LRG mock analysis leads to \emph{unbiased} constraints on $\beta$ (although the linear point now shifts to slightly smaller values, excluding the input value at $>95\%$).

\begin{figure*}
\centering
\includegraphics[width=0.425\textwidth]{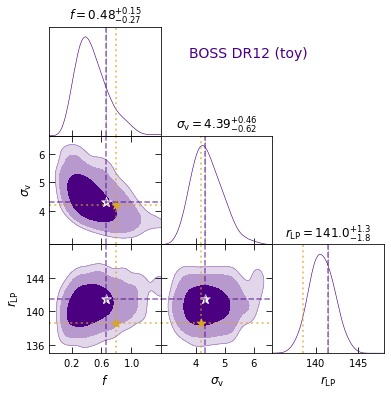}
\hskip 0.1in
\includegraphics[width=0.425\textwidth]{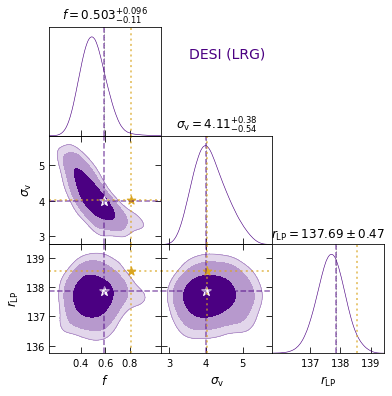}
\caption{Cosmological constraints from the Laguerre reconstruction exercise for the mock BOSS DR12 \emph{(left panel)} and DESI LRG data \emph{(right panel)}. These are alternate versions of the constraints shown in Figs.~\ref{fig:mock_bossdr12_xisLP}-\ref{fig:mock_desilrg_contours}, focusing on the parameters $\{f,\sigv,r_{\rm LP}\}$, formatted identically to Figs.~\ref{fig:mock_bossdr12_contours} and~\ref{fig:mock_desilrg_contours}. The bias in recovering $f$ using the mock DESI LRG data in the \emph{right panel} is discussed in detail in section~\ref{subsec:caveats}.}
\label{fig:contours_subset}
\end{figure*}

\subsubsection{Approximation in equation~\eqref{eq:Sigobs2-model-hack}}
In \eqn{eq:Sigobs2-model}, we introduced an \emph{ad hoc} factor of  $1.38$ to correct the overestimate of $\Sigma_{\rm obs}^2$ by our simple model. The value of this factor very likely depends on the parameters $\{b,\beta,\sigv\}$, although this dependence is probably weak, as seen from the comparison between the DESI LRG and BOSS DR12 configurations above \eqn{eq:Sigobs2-model}. However, the error on the measured value of $\hat\Sigma_{\rm obs}^2$ is only $\sim(1\Mpch)^2$ for the DESI LRGs, which is much smaller than the difference $10.9^2 - 9.3^2 \approx 33 (\Mpch)^2$ between the model prediction without and with the factor. It is possible, therefore, that the MCMC results are sensitive to the precise value of this factor. 

To assess the impact of this approximation, we have repeated the MCMC analysis of the DESI LRG mock data by excluding the measurement of $\Sigma_{\rm obs}^2$. We find that the \emph{a posteriori} distribution of $\beta$ is broader ($\beta=0.220^{+0.044}_{-0.044}$), but has a similar absolute offset of $\sim0.1$ between its median and the input value $\beta=0.334$ as obtained when including $\Sigma_{\rm obs}^2$. The distribution of \sigv, however, now becomes highly biased ($\sigv=1.79^{+0.54}_{-0.54}\,\Mpch$), excluding the input value of $\sigv=4.01\,\Mpch$ at $>99\%$ confidence. The linear point is unaffected, on the other hand, with a constraint $r_{\rm LP} = 138.0^{+0.4}_{-0.4}\,{\rm Mpc}$, which is statistically consistent with the result when including $\Sigma_{\rm obs}^2$ and has a similar uncertainty of $\sim0.3\%$. 

This indicates that the inclusion of $\Sigma_{\rm obs}^2$ along with the approximation in \eqn{eq:Sigobs2-model-hack} actually helps the model perform better with respect to recovering \sigv, while not affecting the constraint on $r_{\rm LP}$. Improving the smearing approximation in \eqns{eq:xi0NL-final}-\eqref{eq:xi4NL-final} is therefore currently more important than modelling the possible parameter dependence of the normalising factor in \eqn{eq:Sigobs2-model-hack}.

\subsubsection{Towards a more accurate model}
Our mocks treat \eqn{eq:xiell-2} as the truth, so the biases in inferred parameters we found arise from the fact that our Zel'dovich smearing approximation is inadequate for describing DESI LRG-like samples, even in this simple case for which differences between model and truth are as small as those shown in the top right hand panel of Fig.~\ref{fig:diagnostic}.  What do we learn from this?

The main deficiency of our model is that its treatment of $\Cal{O}(K^4)$ terms and higher is only approximate.  In principle, we could extend our treatment by organising the higher order corrections as combinations of powers of $K^2$ (i.e., Laplacians in configuration space) and exponentials $\sim\e{-K^2\sigma^2}$ (i.e., Gaussian smoothing kernels) for appropriately chosen smoothing scales $\sigma$. However, in the top left panel of Fig.~\ref{fig:diagnostic}, the zero crossing of $P_2$ is at $k<0.2\hMpc$, while in the simulations of \cite{grieb2016} this does not happen before $k=0.25\hMpc$ (see their fig.~3).  Since their choices of redshift, bias and cosmological model are not very different from ours, this discrepancy almost certainly indicates that mode coupling and/or scale-dependent bias -- which are absent from our mocks -- matter around $k\sim0.2\hMpc$.  Indeed, fig.~2 of \cite{peaksRSD} shows that scale-dependent (density and velocity) bias leave imprints on $\xiell{\ell}{}$, on BAO scales, that are similar in magnitude to the differences between the solid and dashed curves in Fig.~\ref{fig:diagnostic}.  

This suggests that, instead of trying to model \eqn{eq:xiell-2} exactly, we should focus on accounting for mode-coupling and scale-dependent bias. Fortunately, both effects contribute with $k^2$ at leading order, so a `Laplace-Gaussian' expansion of the Laguerre-reconstruction approach should be useful, not so much for mimicking \eqn{eq:xiell-2}, but for including these other effects as well.  We are currently exploring this and will present the results in a forthcoming publication. 

\subsection{Implications for cosmological inference}
\label{subsec:cosmo-inference}
The approach presented above has some interesting implications for cosmological inference in general, which we briefly discuss here.

\subsubsection{Cosmological constraints}
Fig.~\ref{fig:contours_subset} shows the joint constraints on the set of cosmological parameters $\{f,\sigv,r_{\rm LP}\}$ from the two mock data sets. The model recovers the input values of \sigv\ and $r_{\rm LP}$ accurately and precisely for both configurations, and also $f$ for the BOSS DR12 mock.\footnote{The small ($\lesssim0.5\sigma$) difference between the $r_{\rm LP}$ distribution in the \emph{left panel} and that in right panel of Fig.~\ref{fig:mock_bossdr12_xisLP} is due to a slightly different way of handling parameter vectors that don't lead to self-consistent $r_{\rm LP}$ values.} The DESI LRG constraints are biased relative to the input value of the growth rate $f$, for the reasons discussed in section~\ref{subsec:caveats}. Assuming that the Zel'dovich smearing approximation and Laguerre reconstruction scheme can be improved to an accuracy that removes this bias, such a plot can be used to directly compare with \emph{any} cosmological model that predicts the values of these parameters, not only the $\Lambda$CDM model used here. 
For example, it would be interesting to explore the region of parameter space covered by linear theory in this plot, when $\Lambda$CDM parameters such as $\{\Omega_{\rm m},h,\sigma_8\}$ are varied.
This could, in principle, offer a new, model-independent approach to testing dark energy models and/or modified gravity theories.

\subsubsection{Choice of tracers}
In the context of extracting unbiased constraints on $\beta$, it is worth comparing our approach to that of \citet{hamilton92}. In that analysis, one completely ignores the smearing due to non-linear growth to find 
\be
\xiell{2}/[\xiell{0} -{\bar\xi}^{(0)}_{\rm NL}] = \chi_2(\beta)/\chi_0(\beta)\,,
\ee
where the overbar refers to the volume average in \eqn{eq:xibar-def}. In the same approximation, one can also show 
\be
\xiell{4}/[\xiell{2} -(7/5){\bar{\bar\xi}}^{(2)}_{\rm NL}] = \chi_4(\beta)/\chi_2(\beta)\,,
\ee
and 
\be
\xiell{4}/[\xiell{0} - {\bar\xi}^{(0)}_{\rm NL}+ (7/2)({\bar\xi}^{(0)}_{\rm NL}-{\bar{\bar\xi}}^{(0)}_{\rm NL})] = \chi_4(\beta)/\chi_0(\beta)\,, 
\ee
where the double overbar is the volume average in \eqn{eq:xibarbar-def}. Each of these relations could, in principle, be used to extract $\beta$ directly from the observed multipoles of the 2pcf. The MCMC analysis described earlier would then implement this extraction in a fully Bayesian framework. 

In our (more realistic) smearing approximation, this simplicity is lost due to the appearance of different smearing scales \sigeff{\ell}{} in \eqns{eq:xi0NL-final}-\eqref{eq:xi4NL-final}. Equation~\eqref{eq:sigeff-def} shows, however, that the differences between the \sigeff{\ell}{} could be reduced by considering tracers with either large positive $\beta$, or $\beta < 0$. In such cases, \eqns{eq:xi0NL-final}-\eqref{eq:xi4NL-final} can approximately obey Hamilton's relations, potentially leading to substantial gains in cosmological parameter recovery when combined with constraints on $b$ from lensing studies. Galaxies in voids \citep[having $b\sim0$ or $b<0$; see, e.g.,][]{phs18a} are likely to be ideal candidates for realising such samples, and could complement existing efforts to include void statistics in parameter recovery \citep{nadathur+20,nadathur+20-erratum22}. 

Along similar lines, cross-correlating differently biased galaxy sub-samples could also help in parameter recovery, since the corresponding \sigeff{\ell}{} and \beff{\ell}{} values would differ only due to different $b$ values, while having the same values for the cosmological parameters $f$ and \sigv. Such samples would additionally be aided by reduced cosmic variance due to sharing a common volume \citep{mds09,wz20}.

\section{Summary \& Conclusion}
\label{sec:conclude}
We have charted a path towards using the observed, anisotropic, large-scale 2-point correlation function (2pcf) of galaxies in redshift space to extract cosmological information without relying on a fiducial cosmological model. Such a programme is of particular relevance for testing generic cosmological models, both within general relativity and beyond.
Our framework assumes that non-linear growth leads to a smearing of the BAO feature by an approximately Gaussian kernel, and relies on the `Laguerre reconstruction' of the baryon acoustic oscillation (BAO) feature -- especially its linear point $r_{\rm LP}$ \citep{LP2016} -- introduced by \citet{nsz21a}. We have substantially extended the Laguerre reconstruction framework so as to simultaneously model the redshift space multipoles $\xiell{\ell}(s)$ of the observed 2pcf of biased tracers, for $\ell=0,2,4$.

The starting point of our framework is the Zel'dovich smearing approximation discussed in section~\ref{sec:streaming}. We showed that, in the range $55\lesssim s/(\Mpch)\lesssim125$, \emph{all} three multipoles under this approximation can be reduced to the \emph{isotropic} Gaussian smearing of the linear theory real space 2pcf, but with a smoothing scale \sigeff{\ell}{} that depends on $\ell$ (equation~\ref{eq:sigeff-def}). The accuracy of this approximation is sufficient for use with existing data sets such as the BOSS DR12 LRG sample, but is expected not to be accurate enough for upcoming samples such as DESI LRGs (sections~\ref{subsec:modelaccuracy} and~\ref{subsec:caveats}).  

The Laguerre reconstruction scheme in real space relies on a polynomial approximation for the linear theory 2pcf (equation~\ref{eq:xilin-poly}), which becomes an expansion in generalised Laguerre functions after a Gaussian smearing \citep{nsz21a,nsz21b,nsz22}. In the approximation mentioned above, this feature extends to \emph{all} the redshift space multipoles \xiell{\ell} for $\ell=0,2,4$ (section~\ref{sec:recon}). While this result was already noticed for the monopole $\ell=0$ by \citet{nsz21b} and used by those authors as well as \citet[][PS22]{ps22}, our extension of the same ideas to $\ell=2$ and $4$ has significant consequences for parameter recovery. Namely, since  the \emph{same} polynomial coefficients appear in the description of all three multipoles, the additional dependence of these observables on the linear bias $b$, the modified growth rate $\beta=f/b$ and the linear theory velocity dispersion \sigv\ can be leveraged to constrain these parameters from observations of the \xiell{\ell}. We emphasize that, unlike standard cosmological BAO analyses \citep[e.g.,][]{anderson+14,cuesta+16,gil-marin+20}, as well as recent work on using the linear point for cosmological inference \citep{hzs23}, our approach \emph{does not} assume a fiducial cosmology to model the shape of the BAO feature in redshift space, instead relying on an agnostic basis of functions (here, polynomials).\footnote{Of course, converting angles and redshifts to distances does require a model, but this plays the same role as the standardisation step in the MCMC analysis described in section~\ref{sec:analysis}. 
This can be easily dealt with by either always focusing on scaled quantities such as $y_{\rm LP}\equiv r_{\rm LP}/D_{\rm V}$ \citep{LP2016}, and similarly for \sigv, or always performing comparisons between model predictions and observational constraints for some length scale by multiplying the former by $D_{\rm V,fid}/D_{\rm V,model}$ \citepalias{ps22}, where $D_{\rm V}$ is the volume-averaged distance scale in equation~(6) of \citet{cuesta+16} and the subscripts `fid' and `model' refer to, respectively, the fiducial cosmology used for distance conversions in the observational analysis and the cosmology used in the model prediction. Alternatively, to avoid converting angles and redshifts into comoving distances, some analyses work exclusively with the angular correlation function $\omega(\theta)$ in a relatively narrow redshift slice \citep{wthetaBAO, wtheta2020, wtheta2022}, or the one dimensional correlation $\xi(\Delta z)$ along the line of sight \citep{xizBAO}.  Since the BAO feature in these observables would also be smeared, they too must be reconstructed, so we are in the process of extending our Laguerre methodology to treat such observables as well.}

We tested our framework using mock observations that mimic the existing BOSS DR12 LRG and expected DESI LRG samples (see Table~\ref{tab:toy_samples}). The construction of the mock observations and our MCMC analysis setup are described in section~\ref{sec:analysis}, with results presented in section~\ref{sec:results}. As anticipated, the reconstruction works well for the BOSS DR12 LRG configuration, with unbiased results for the cosmological parameters $\beta$, \sigv\ and the linear point $r_{\rm LP}$. For the DESI LRG configuration, while the recovery of \sigv\ and $r_{\rm LP}$ remains unbiased, the constraint on $\beta$ excludes the `true' input value at $>99\%$ confidence. The reason for this failure is closely connected to the breakdown of the smearing approximation mentioned above, and is discussed in section~\ref{subsec:caveats}. We also commented in that section on some ways forward in improving the approximation, as well as the role of a reasonably tight prior on the value of $b$, which should be achievable using cross-correlations of the galaxy sample with weak gravitational lensing observations.

Finally, we have not exploited the fact that several of the eigen-coefficients describing the real space linear theory 2pcf $\xi_{\rm lin}(r)$ are statistically consistent with zero in the \emph{a posteriori} distributions of both the mock samples we studied (Figs.~\ref{fig:mock_bossdr12_contours} and~\ref{fig:mock_desilrg_contours}). This fact, which would allow us to substantially decrease the parameter space by only focusing on the most important principle components in the space of polynomial coefficients, can already be determined in the initial, linear Gaussian step of the analysis. It should thus be straightforward to include a step where we only retain these components in the MCMC analysis. Intuitively, these surviving components are likely to be associated with odd, rather than even, Laguerre functions \citep[see, e.g., the discussion in][]{nsz21a}. We will test and incorporate this idea, which should further decrease the uncertainties in the recovered parameters, in our future applications.

Our results, and the discussion above, show that it is feasible to extract meaningful constraints on the cosmological parameters $\{f,\sigv,r_{\rm LP}\}$ from upcoming surveys (Fig.~\ref{fig:contours_subset}), \emph{without} explicitly assuming an underlying cosmological model such as $\Lambda$CDM, and that the sensitivity of this approach can be enhanced by careful choices of sample selection (section~\ref{subsec:cosmo-inference}). 
This bodes well for comparison exercises with a wider class of cosmological models, such as those incorporating dark energy or departures from general relativity. A key requirement for such an exercise would be the prediction of parameters equivalent to $\{f,\sigv,r_{\rm LP}\}$ in these alternate cosmological models. We will expand on this theme in future work.

 
\section*{Acknowledgments}
The research of AP is supported by the Associateship Scheme of ICTP, Trieste.
This work made extensive use of the open source computing packages NumPy \citep{vanderwalt-numpy},\footnote{\url{http://www.numpy.org}} SciPy \citep{scipy},\footnote{\url{http://www.scipy.org}} Matplotlib \citep{hunter07_matplotlib},\footnote{\url{https://matplotlib.org/}} and Jupyter Notebook.\footnote{\url{https://jupyter.org}}

\section*{Data availability}
The MCMC chains produced in this work will be made available upon reasonable request to the authors.

\bibliography{references}
 
\appendix
\section{}
\subsection{Details of the Zel'dovich smearing approximation}
\label{app:streamingapprox}
Here we provide some details of the manipulations involved in deriving the final expressions \eqref{eq:xi0NL-final}-\eqref{eq:xi4NL-final} for the Zel'dovich smearing approximation to \eqn{eq:xiell-2} for \xiell{\ell}.

We use the differential Bessel identity \citep[see 10.1.24 of][]{abramowitz-stegun}
\be
j_\ell(x) = (-1)^\ell\,x^\ell\,\left(\frac1x\frac{\der}{\der x}\right)^\ell\,j_0(x)\,,;\quad\ell=1,2,\ldots\,,
\label{eq:diffBessel}
\ee
and the Fourier association $\nabla^2\leftrightarrow-k^2$ to write \eqref{eq:xiell-2} as
\begin{align}
\xiell{\ell}(s) &= \beff{\ell}{2}\,i^\ell\,s^\ell\left(\frac{1}{s}\frac{\p}{\p s}\right)^\ell(-\nabla^{-2})^{\ell/2}\,\xi_0(s|\sigeff{\ell}{})\,,
\label{eq:xiell-3}
\end{align}
where we defined $\xi_0(s|\sigma)$ in \eqn{eq:xi0(s|sigma)}.

The integro-derivative operator acting on $\xi_0(s|\sigeff{\ell}{})$ can be simplified for $\ell=2$ and $4$ along the lines of \citet{hamilton92}. First, note that the 3-dimensional inverse Laplacian of any spherically symmetric function $\xi(s)$ is \citep[][section 2.4]{binney-tremaine-GalDyn}
\be
\nabla^{-2}\xi(s) = -\frac{1}{s}\int_0^s\,\der u\,u^2\,\xi(u) - \int_s^\infty\der u\,u\,\xi(u)\,.
\label{eq:invLaplacian}
\ee
Applying the inverse Laplacian once more and exchanging the order of the resulting double integral gives, after some straightforward algebra,
\begin{align}
\nabla^{-4}\xi(s) &= -\frac{1}{6s}\int_0^s\der u\,u^4\,\xi(u) - \frac{s}{2}\int_0^s\der u\,u^2\,\xi(u) \notag\\
&\ph{\frac16\int\der u}
-\frac{1}{2}\int_s^\infty\der u\,u^3\,\xi(u) -\frac{s^2}{6}\int_s^\infty\der u\,u\,\xi(u)\,.
\end{align}
Finally, define the volume averages $\bar{\xi}(s)$ and $\bar{\bar{\xi}}(s)$ of any function $\xi(s)$ as
\begin{align}
\bar{\xi}(s) &\equiv \frac{3}{s^3}\int_0^s\der u\,u^2\,\xi(u)\,,
\label{eq:xibar-def}\\
\bar{\bar{\xi}}(s) &\equiv \frac{5}{s^5}\int_0^s\der u\,u^4\,\xi(u)\,,
\label{eq:xibarbar-def}
\end{align}
Repeated differentiation of $\nabla^{-\ell}\xi_0(s|\sigeff{\ell}{})$ for $\ell=0,2$ and $4$ then gives us \eqns{eq:xi0NL-final}-\eqref{eq:xi2NL-final}.

The following identities are useful for the numerical evaluation of the approximation shown in the  \emph{right panel} of Fig.~\ref{fig:diagnostic},
\begin{align}
\xi_0(s|\sigma) &= \int\der\ln k\,\Delta_{\rm lin}^2(k)\,\e{-k^2\sigma^2/2}\,j_0(ks)\,, \notag\\
\bar{\xi}_0(s|\sigma) &= \int\der\ln k\,\Delta_{\rm lin}^2(k)\,\e{-k^2\sigma^2/2}\,3j_1(ks)/(ks)\,,\notag\\
\bar{\bar{\xi}}_0(s|\sigma) &= \int\der\ln k\,\Delta_{\rm lin}^2(k)\,\e{-k^2\sigma^2/2}\,\frac{5}{(ks)^3}\bigg[2\sin(ks) \notag\\
&\ph{\int\der\ln k\,\Delta_{\rm lin}^2(k)\,\e{-k^2}}
+ \left((ks)^2-6\right)j_1(ks)\bigg]\,.
\label{eq:xi0bars-integrals}
\end{align}

\subsection{Gauss-Poisson covariance matrix}\label{sec:Cgp}
Here, we discuss some implications of the Gauss-Poisson approximation \eqref{eq:C_ell1ell2(si,sj)} for the covariance of 2pcf measurements that are relevant for understanding the magnitude of errors on measurements of multipoles with increasing $\ell$. In the following, we focus on Fourier space and denote $\mu_k\equiv\hat n\cdot\hat k$ as $\mu$ for brevity.

As discussed in the main text, at BAO scales, the smearing approximation for the non-linear power spectrum works well. We can then write the anisotropic power spectrum as
\begin{equation}
P(k,\mu)\approx b^2\,(1 + \beta\mu^2)^2\,P_{\rm lin}(k)\,
{\rm e}^{-k^2\sigma_{\rm v}^2(1-\mu^2)}{\rm e}^{-k^2\sigma_{\rm v}^2\mu^2(1+f)^2} ,
\end{equation}
for which the integral over $\mu$ which defines $\sigma^2_{\ell_1\ell_2}$ in \eqn{eq:sig2_ell1ell2(k)} can be done analytically for all $(\ell_1,\ell_2)$.  This is because the integral over $\mu$ is simplified by noting that if 
\begin{equation}
{\cal I}_{j}(\kappa) \equiv \int_{-1}^1 \frac{d\mu}{2}\,{\rm e}^{-\kappa^2\mu^2/2} \, \mu^{j} ,
\end{equation}
then 
\begin{equation}
\kappa^2\, {\cal I}_{2(n+1)}(\kappa) 
= (2n+1)\,{\cal I}_{2n}(\kappa) - {\rm e}^{-\kappa^2/2} , 
\end{equation}
with 
\begin{equation}    
{\cal I}_{0}(\kappa) = \frac{\sqrt{\pi}}{2}\, \frac{{\rm erf}(\kappa/\sqrt{2})}{\kappa/\sqrt{2}} .
\end{equation}
(Note that all ${\cal I}_{2n+1}(\kappa)=0$.)
The multipoles $P_0$ and $P_2$ of $P(k,\mu)$ are then
\begin{equation}
P_0(k) 
= b^2\,P_{\rm lin}(k)\,{\rm e}^{-k^2\sigma_{\rm v}^2}
  \sum_{j=0}^2 {2\choose j} \beta^j \,{\cal I}_{2j}(K),
\end{equation}
and
\begin{equation}
\frac{P_2(k)}{5} 
= b^2\,P_{\rm lin}(k)\,{\rm e}^{-k^2\sigma_{\rm v}^2}
  \sum_{j=0}^2 {2\choose j} \beta^j \,
  \left[\frac{3{\cal I}_{2j+2}(K) - {\cal I}_{2j}(K)}{2} \right] \, ,
\end{equation}
and $\sigma_{00}^2$ becomes
\begin{align}
\frac{\sigma^2_{00}}{2V_{\rm sur}/({\bar n}V_{\rm sur})^2} 
&= 1 + 2\,{\bar n}P_0(k) \notag\\
&+ b^4\,{\bar n}^2P^2_{\rm lin}(k)\,
  {\rm e}^{-2k^2\sigma_{\rm v}^2}
\sum_{j=0}^4 {4\choose j} \, \beta^j \,{\cal I}_{2j}(\sqrt{2}K)
\end{align}
where $K^2 = 2k^2\sigma_{\rm v}^2 f(2+f)$. The other $\ell_1\ell_2$ pairs simply involve more powers of $\mu^2$, so yield other combinations of the ${\cal I}_{2j}$.

This leaves the integral over $k$ as the only one which must be done numerically, since only the shot-noise piece is analytic:  
\begin{align}
&\int \frac{dk\,k^2}{2\pi^2}\,\bar{j}_{\ell_1}(ks_i)\,\bar{j}_{\ell_2}(ks_j)\notag\\
&= \frac{(4\pi)^2}{V_i V_j}\int ds\,s^2\,W_i(s)\int dt\,t^2\,W_j(t)\int  \frac{dk\,k^2}{2\pi^2}\,j_{\ell_1}(ks)\,j_{\ell_2}(kt)\,\nonumber\\
&= \frac{4\pi}{V_i V_j}\int ds\,s^2\,W_i(s)\,W_j(s)\, .
\end{align}
For non-overlapping tophat bins of unit height, the integral is zero unless $i=j$, in which case the expression above becomes $1/V_i$.  I.e., this shotnoise term contributes as $\delta_{ij}\,2\,(V_{\rm sur}/V_i)/({\bar n}V_{\rm sur})^2$, where we recognize that ${\bar n}V_{\rm sur}=N$, so $2\,(V_{\rm sur}/V_i)/N^2\approx (V_i/V_{\rm sur})^{-1}/{N\choose 2}$.  This diverges in the limit of small bins, as expected.  For typical BAO samples which have ${\bar n} P\sim 1$ at $k\sim 0.1\,\hMpc$, the other terms dominate unless the bins are substantially narrower than a tenth of an Mpc.  

However, to gain intuition, it is useful to invoke the smearing approximation and write the anisotropic power spectrum as 
\begin{equation}
P(k,\mu)\approx \sum_\ell P_\ell(k)\,\Pell{\ell}(\mu)
\end{equation}
with 
\begin{equation}
P_\ell(k) \equiv P_{\rm lin}(k)\, \beff{\ell}{2}\,\e{-k^2\sigeff{\ell}{2}/2}\,,
\end{equation}
where \beff{\ell}{} and \sigeff{\ell}{} were defined in \eqns{eq:beff-def} and~\eqref{eq:sigeff-def}, respectively.
Then, the orthogonality of Legendre polynomials means that the $1/\bar{n}^2$ term in \eqn{eq:sig2_ell1ell2(k)} gives $\delta_{\ell_1 \ell_2}\,(2\ell_1 +1)\,2/(\bar{n}^2V_{\rm sur})$.  The term proportional to $(2P(k,\mu)/\bar{n}V_{\rm sur})$ involves the product of three Legendre polynomials, which equation~(\ref{eq:Legendre-3j}) simplifies to 
\begin{equation}
\frac{2}{\bar{n}V_{\rm sur}}\,(2\ell_1+1)(2\ell_2+1)\,2 \sum_{\ell} \tjred{\ell_1}{\ell_2}{\ell}^2 \,P_\ell(k)\,.
\end{equation}
Finally, the term proportional to $P^2(k,\mu)$ involves four Legendre polynomials.  It can be simplified by using the fact that 
\begin{equation}
 \Pell{\ell_1}(\mu)\,\Pell{\ell_2}(\mu) =  \sum_{\ell = |\ell_1-\ell_2|}^{\ell_1+\ell_2}  \tjred{\ell_1}{\ell_2}{\ell}^2 (2\ell + 1)\,\Pell{\ell}(\mu);
\end{equation}
this reduces the expression to the product of three polynomials, and then equation~(\ref{eq:Legendre-3j}) further simplifies it to 
\begin{align}
\frac{2}{V_{\rm sur}}&\,(2\ell_1+1)(2\ell_2+1)
\sum_{\ell = |\ell_1-\ell_2|}^{\ell_1+\ell_2} \tjred{\ell_1}{\ell_2}{\ell}^2 \,(2\ell + 1) \notag\\
&\times 
\sum_{\ell^\prime} \sum_{\ell^{\prime\prime}} \tjred{\ell}{\ell^\prime}{\ell^{\prime\prime}}^2 P_{\ell^\prime}(k)\, P_{\ell^{\prime\prime}}(k) .
\end{align}
When $\ell_1=\ell_2=0$ (for the covariance of the monopole) then this term becomes 
\begin{equation}
\frac{2}{V_{\rm sur}}\,
 \sum_{\ell^\prime} \sum_{\ell^{\prime\prime}} \tjred{0}{\ell^\prime}{\ell^{\prime\prime}}^2 
 P_{\ell^\prime}(k)\, P_{\ell^{\prime\prime}}(k) 
= \frac{2}{V_{\rm sur}}\,\sum_{\ell} \frac{P_\ell^2(k)}{2\ell + 1},
\end{equation}
making 
\begin{align}
\sigma^2_{00}(k) &= \frac{2}{V_{\rm sur}}\,\sum_{\ell} \frac{P_\ell^2(k)}{2\ell + 1} 
+ \frac{2}{\bar{n}V_{\rm sur}}\,2P_0(k) + \frac{2}{\bar{n}^2V_{\rm sur}} \notag\\
&= \frac{2}{V_{\rm sur}}\,\left[\left(P_0(k) + \frac{1}{\bar{n}}\right)^2 +\sum_{\ell > 0} \frac{P_\ell^2(k)}{2\ell + 1} \right]. 
\end{align}
Notice that this has a contribution from the higher order multipoles, in addition to the naive term one might expect from the monopole itself.  Similarly,
\begin{align}
\sigma^2_{22}(k) &= \frac{2}{V_{\rm sur}} \times 5\,\bigg[\left(P_0(k) + \frac{1}{\bar n}\right)^2 + 
\frac{3}{7}\,P^2_2(k) \notag\\
&\ph{2}
+ \frac{1789}{9009}\,P_4^2(k) + \frac{4}{7} \left(P_0(k) + \frac{1}{\bar n}\right)[P_2(k) + P_4(k)] \notag\\
&\ph{2V_{\rm sur}}
+ \frac{24}{77} P_2(k)P_4(k)  
\bigg]
\end{align}
and
\begin{align}
\sigma^2_{44}(k) &= \frac{2}{V_{\rm sur}} \times9\,\bigg[\left(P_0(k) + \frac{1}{\bar n}\right)^2 + 
\frac{1789}{5005}\,P^2_2(k) \notag\\
&\ph{2}
+ \frac{529}{1001}\frac{9}{17}\,P_4^2(k) 
+ \frac{360}{1001} P_2(k)P_4(k)  \notag\\
&\ph{2}
+ \left(P_0(k) + \frac{1}{\bar n}\right) \left(\frac{40}{77} P_2(k) + \frac{324}{1001}P_4(k)\right) 
\bigg].
\end{align}
The cross terms are 
\begin{align}
\sigma^2_{02}(k) &= \frac{2}{V_{\rm sur}}\bigg[\,2\left(P_0(k) + \frac{1}{\bar n}\right)\,P_2(k) \notag\\
&+ \frac{2}{7}\left(P^2_{2}(k) + 2P_{2}(k) P_{4}(k) + \frac{50}{99}P^2_{4}(k) \right)\bigg]\,,\\
\sigma^2_{04}(k) &= \frac{2}{V_{\rm sur}}\bigg[2\left(P_0(k) + \frac{1}{\bar n}\right)\,P_4(k) \notag\\
&+ \frac{2}{7}\left(\frac{9}{5}P^2_{2}(k) + \frac{81}{143}P^2_{4}(k) + \frac{20}{11}P_2(k)\,P_4(k) \right)\bigg]
\end{align}
and 
\begin{align}
\sigma^2_{24}(k) &= \frac{2}{V_{\rm sur}}\bigg[2P_2(k) P_4(k) \notag\\
&\ph{2/Vs}
+ \frac{2}{7}
\left(P_0(k) + \frac{1}{\bar n}\right) \left(18P_2(k) + \frac{100}{11}P_4(k)\right) \notag\\
&+ \frac{2}{7}\left(\frac{54}{11}P^2_{2}(k) + \frac{450}{143}P^2_{4}(k) + \frac{788}{143}P_2(k) P_4(k) \right)\bigg].
\end{align}
For most galaxy samples $P_0(k) > P_2(k) > P_4(k)$ around BAO scales $k\sim0.1\,\hMpc$ \cite[see, e.g., the \emph{left panel} of Fig.~\ref{fig:diagnostic}, or fig.~3 of][]{grieb2016}, which suggests that 
\begin{equation}
    \sigma^2_{\ell\ell}(k)\approx (2\ell + 1)\,\sigma^2_{00}(k)
\end{equation}
should be reasonably accurate.  Indeed, this expression mimics the dependence on $\ell$ shown in fig.~4 of \cite{grieb2016} quite well.  Similarly, 
\begin{equation}
    \sigma^2_{0\ell}(k)\approx \frac{2}{V_{\rm sur}}\,2\left[P_0(k) + \frac{1}{\bar n}\right]\,P_\ell(k) .
\end{equation}
In this case, 
\begin{equation}
    R_{\ell_1\ell_2}(k) \equiv \frac{\sigma^2_{\ell_1\ell_2}(k)}{\sigma_{\ell_1\ell_1}(k)\sigma_{\ell_2\ell_2}(k)}
    \approx \frac{\sigma^2_{\ell_1\ell_2}(k)/\sigma^2_{00}(k)}{\sqrt{(2\ell_1 + 1)(2\ell_2 + 1)}} .
\end{equation}
making 
\begin{equation}
    R_{0\ell}(k)\approx 
    \frac{\sigma^2_{0\ell}(k)/\sigma^2_{00}(k)}{\sqrt{(2\ell + 1)}}
    = \frac{2}{\sqrt{2\ell + 1}}\frac{\bar{n}P_\ell(k)}{\bar{n}P_0(k) + 1}.
\end{equation}
This seems to be in good agreement with the $k_i=k_j$ (diagonal) elements shown in fig.~5 of \cite{grieb2016}.

\label{lastpage}

\end{document}